\documentclass{article}

\usepackage[preprint]{neurips_2026}

\usepackage[utf8]{inputenc} 
\usepackage[T1]{fontenc}    
\usepackage{hyperref}       
\usepackage{url}            
\usepackage{booktabs}       
\usepackage{amsfonts}       
\usepackage{nicefrac}       
\usepackage{microtype}      
\usepackage{xcolor}         
\usepackage{amsmath}
\usepackage{graphicx}
\usepackage{booktabs} 
\usepackage{multirow}
\usepackage{xspace}

\usepackage{color}
\usepackage{placeins}

\usepackage[table]{xcolor}
\usepackage{multirow}
\newcommand{\ns}{\texttt{Navier\_Stokes}\xspace}
\newcommand{\react}{\texttt{Diff\_Reaction}\xspace}
\newcommand{\shallow}{\texttt{Shallow\_Water}\xspace}
\newcommand{\actmatter}{\texttt{Active\_Matter}\xspace}
\newcommand{\MHD}{\texttt{MHD\_64}\xspace}
\newcommand{\trltwo}{\texttt{TRL\_2D}\xspace}
\newcommand{\cartesiantwoD}{($x,y$)}
\newcommand{\cartesianthreeD}{($x,y,z$)}
\usepackage{textcomp}

\DeclareSymbolFont{matha}{OML}{txmi}{m}{it} 
\DeclareMathSymbol{\varv}{\mathord}{matha}{118}

\title{Adaptive Coordinate Transforms for Neural Operators}

\author{
Chaoyu Liu$^{1}$$^,$\thanks{Equal contribution. Correspondence to: \texttt{cl920@cam.ac.uk}} \quad
Zhonghao Li$^{2}$$^,$\footnotemark[1] \quad
Gaohang Chen$^{3}$ \quad
Zakhar Shumaylov$^{1}$ \AND
Zhongying Deng$^{1,4}$ \quad
Qian Zhang$^{2}$ \quad
Zhonghua Qiao$^{3}$ \quad
Carola-Bibiane Schönlieb$^{1}$ \\
\\
$^1$ Department of Applied Mathematics and Theoretical Physics, University of Cambridge \\
$^2$ School of Science, Harbin Institute of Technology, Shenzhen \\
$^3$ Department of Applied Mathematics, The Hong Kong Polytechnic University \\
$^4$ Current affiliation: Department of Radiology, University of Cambridge
}

\begin{document}

\maketitle


\begin{abstract}
Neural operators have achieved promising performance on partial differential equations (PDEs), but most existing models are built on fixed Eulerian coordinates. This mismatch between evolving physical structures and static coordinates creates spatial misalignment, leading to unnecessarily non-local operator mappings and reinforcing a smoothness preference near sharp transitions. Inspired by adaptive coordinate transformations in classical PDE analysis, we propose the Adaptive Coordinate Transform (ACT) block, a plug-and-play module for data-driven geometric adaptation in neural operators. ACT blocks resolve this structural limitation by learning adaptive coordinate systems within the operator learning pipeline. Specifically, given an input feature, the ACT block learns a coordinate transformation and represents the same feature under the transformed coordinates via differentiable sampling. This operation preserves the underlying signal while changing its spatial representation, equivalent to expressing the same physical quantity in different coordinate systems. By adapting the coordinate system to the data, ACT allows the network to better track evolving structures, reduce operator complexity, and dynamically focus on critical features to improve learning. We evaluate the proposed approach across diverse PDE benchmarks and multiple neural operator architectures. Experimental results demonstrate consistent and significant improvements in predictive accuracy, indicating that learning coordinate systems provides a powerful mechanism for enhancing operator learning.
\end{abstract}
\section{Introduction}

Neural operators have emerged as a powerful framework for solving partial differential equations (PDEs) \citep{kovachki2023neural}. By learning mappings between input and output fields, they can serve as efficient surrogates for classical numerical solvers. Architectures such as Fourier Neural Operators (FNO) \citep{li2021fourier}, ConvNeXt-UNet (CNextU) \citep{convunet}, and Transolver \citep{wu2024transolverfasttransformersolver} have shown strong performance in modeling complex physical dynamics.

Despite their success, a fundamental limitation exists: most existing neural operators are built upon fixed Eulerian coordinates. This static coordinate system is inherently misaligned with the evolving physical structures (e.g., propagating waves, moving vortices) that characterize many PDEs. This \emph{spatial misalignment} leads to highly non-local and complex input-output mappings, significantly increasing learning complexity and hindering generalization \citep{hu2024better}. Furthermore, these operators exhibit a secondary challenge: a \emph{smoothness preference} \citep{rahaman2019spectral}. Constrained by their inherent inductive biases, they tend to favor overly smooth mappings and struggle to maintain the correct regularity of the solution, often blurring high-frequency physical phenomena like shocks or sharp transitions.

From the perspective of classical PDE analysis, these challenges of complexity and regularity are primarily consequences of using fixed Eulerian coordinates. Rather than solving such evolving dynamics within a static coordinate system, classical methods often employ adaptive coordinate transformations, ranging from Arbitrary Lagrangian-Eulerian (ALE) formulations \citep{hirt1974arbitrary} to moving mesh methods \citep{huang2010adaptive}, to dynamically reshape the computational coordinates. By allowing the coordinate system to track advecting structures and locally compress around sharp gradients, these techniques simultaneously reduce spatial misalignment and avoid excessive numerical smoothing. Consequently, they unravel complex spatiotemporal evolution into a smoother, more localized, and better resolved form.

Inspired by these classical principles of adaptive coordinate transformation, we propose the Adaptive Coordinate Transform (ACT) block, a plug-and-play module designed to equip neural operators with data-driven geometric realignment capabilities. By predicting a coordinate adjustment field from the current feature map, the ACT block resamples intermediate features via differentiable sampling and integrates them through a residual connection. Specifically, we interleave ACT blocks layer-wise within foundational operator backbones (e.g., FNO, CNextU, and Transolver), creating an architecture capable of continuous geometric realignment.

Crucially, we uncover an autonomous role decoupling mechanism within this layer-wise architecture. Based on their topological locations, the ACT blocks dynamically divide their labor to mimic classical adaptive techniques. Intermediate blocks act as progressive spatial trackers, realigning evolving features to neutralize spatial misalignment and maintain a stable latent space. Conversely, the final ACT block acts as a regularity modulator. By implicitly compressing the local coordinate space, it leverages the Jacobian multiplier to explicitly amplify spatial gradients, effectively counteracting the smoothness preference and recovering sharp physical transitions.

In summary, our work bridges the gap between classical adaptive mesh techniques and modern operator learning. The main contributions are as follows:

\begin{itemize}
\item We propose the Adaptive Coordinate Transform (ACT) block, a learnable geometric realignment module inspired by classic PDE techniques, specifically designed to overcome spatial misalignment and the smoothness preference of neural operators built on fixed Eulerian coordinates.
\item Extensive experiments across diverse foundational architectures (FNO, CNextU, and Transolver) demonstrate that ACT-augmented models significantly improve predictive accuracy and better capture complex dynamics.
\item We reveal an autonomous role decoupling phenomenon in layer-wise networks: intermediate ACTs track advection for latent alignment, while the terminal ACT reconstructs sharp gradients via Jacobian-driven spatial compression.
\end{itemize}

\section{Related Work}

\paragraph{Neural Operators on Fixed Coordinates}
Neural operators learn mappings between function spaces and have achieved strong performance on PDEs. Representative paradigms include spectral operators such as the Fourier Neural Operator (FNO) \citep{li2021fourier}, encoder-decoder baselines built on modern convolutional blocks \citep{convunet}, and transformer-based architectures such as Transolver \citep{wu2024transolverfasttransformersolver}. These models offer distinct inductive biases for spectral mixing, locality-aware hierarchical feature extraction, and long-range interaction modeling, respectively. However, these approaches generally assume a fixed Eulerian coordinate system and do not explicitly model coordinate transformations. For dynamics involving transport and advection, they address spatial misalignment only indirectly through spectral global coupling, stacked local convolutions, or attention-based global interactions, so the operator remains unnecessarily non-local in the original coordinates.

\paragraph{Coordinate Adaptation in PDE Solvers and PDE Learning}
Reducing complexity through adaptive coordinate transformations or mesh adaptation is standard in PDE computation. In learning-based settings, some works learn deformation maps or adaptive meshes to improve discretization and solver efficiency \citep{rowbottom2025gadaptivity,zhang2024towards,hu2024better}, while others use learned mappings to parameterize irregular domains or canonicalize data before operator learning \citep{li2023fourier,shumaylov2025lie}. These approaches are related in spirit, but they mainly target mesh optimization, domain parameterization, or preprocessing rather than layer-wise geometric realignment of latent operator representations.

\paragraph{Learnable Deformation Modules}
Learnable deformation is also well established in deep learning, especially in image registration \citep{de2017end,balakrishnan2019voxelmorph}, deformable convolution \citep{dai2017deformableCNN}, and deformable attention \citep{zhu2020deformableTrans}. These methods motivate ACT architecturally, especially through local offset prediction and differentiable sampling. However, they are mostly designed for vision tasks, and even deformation-based neural operator variants outside PDE learning \citep{mothish2025dnod} remain application-specific. In contrast, we treat adaptive coordinate transformation as a native, layer-wise component of neural operator learning, using geometric realignment throughout the backbone to reduce spatial misalignment and simplify the learned operator.
\section{Method}  \label{sec:method}

This section presents our framework for learning neural operators under adaptive coordinate transformations. We first describe the spatial misalignment induced by fixed Eulerian coordinates, then introduce the proposed ACT block, and finally show how it is integrated with a neural operator backbone.

\subsection{Problem Setup}

Let $\Omega \subset \mathbb{R}^d$ be a bounded spatial domain equipped with a fixed Eulerian coordinate system. Let $U(\Omega \times [0, T]; \mathbb{R}^{d_u})$ and $V(\Omega \times [0, T]; \mathbb{R}^{d_v})$ denote the input and output function spaces, respectively. We consider an operator $\mathcal{F}$ that maps functions from $U$ to $V$ as:
\begin{equation} 
\mathcal{F}: U \rightarrow V, \quad u(x,t) \mapsto v(x,t) = \mathcal{F}[u](x,t),
\end{equation}
which is typically induced by an underlying PDE. Here $u \in U$ and $v \in V$ are spatiotemporal functions evaluated at a given point $x \in \Omega$ and time $t \in [0, T]$.
Depending on the task, $u(x,t)$ may represent the initial and boundary conditions or a forcing term, while $v(x,t)$ typically represents the solution field at a future time step.

Neural operators approximate $\mathcal{F}$ by a parameterized network $\mathcal{F}_\theta$, which is typically formulated as:
\begin{equation} \label{eq:operator_mapping}
\mathcal{F}_\theta = \mathcal{Q} \circ \mathcal{G}^{L-1}_\theta \circ \cdots \circ \mathcal{G}^{0}_\theta \circ \mathcal{P},
\end{equation}
where $\mathcal{P}$ pointwise lifts the input function to a latent feature space, $\mathcal{G}^{l}_\theta$ denotes the $l$-th latent operator update (e.g., a spatial integral operator using Fourier or attention mechanisms) acting on the entire domain, and $\mathcal{Q}$ is a local projection mapping the final hidden representation back to the physical output space at $x$.

\subsection{Spatial Misalignment and Geometric Realignment}


The main difficulty arises from the mismatch between evolving physical structures and fixed Eulerian coordinates. 
Physical dynamics are inherently characterized by continuous transport and diffusion, causing structures (e.g., a propagating wave or a moving vortex) to naturally advect and spread across the spatial domain over time.

When such dynamics are modeled on a fixed Eulerian coordinate system, the fundamental mismatch between the static coordinates and evolving structures breaks this spatial alignment, resulting in a highly non-local and complex input-output mapping. 
For the neural operator $\mathcal{F}_\theta: u(x,t) \mapsto v(x,t)$ in Eq.~\eqref{eq:operator_mapping}, this means that a large part of the apparent complexity may come from spatial misalignment in the representation rather than from the underlying dynamics themselves.

However, classical PDE analysis provides a clear solution to this problem: rather than solving such evolving dynamics within a static coordinate system, classical methods often employ coordinate transformations to simplify the problem. 
Concretely, we can define a coordinate transformation 
\begin{equation}
    \psi: \Omega \rightarrow \Omega, \quad x \mapsto \xi := \psi(x),
\end{equation}
that maps the original Eulerian coordinates to a new coordinate system that inherently tracks the evolving structures. Then the neural operator can be reformulated under the new coordinates as:
\begin{equation}
\bar{\mathcal{F}}_\theta: \bar u(\xi,t) \mapsto \bar v(\xi,t),
\end{equation}
where $\bar u(\xi,t) = u\left(\phi(\xi),t\right), \bar v(\xi,t) = v\left(\phi(\xi),t\right)$ and $\phi := \psi^{-1}$ denotes the inverse map. Therefore, the transformed operator $\bar{\mathcal{F}}_\theta$ may admit a simpler representation, since adaptive coordinate transformations can reduce spatial misalignment and reorganize complex spatiotemporal evolution into a smoother and more localized form \citep{hirt1974arbitrary,huang2010adaptive}.


\subsection{Adaptive Coordinate Transform (ACT) Block}

These observations suggest that geometric realignment should be handled explicitly rather than absorbed implicitly by the operator backbone. Following this idea, we introduce an adaptive coordinate transform (ACT) block, which uses the current feature map to realign sampling locations before the next operator update.
Motivated by this perspective, ACT realizes geometric realignment in a learnable, layer-wise manner through feature-conditioned local sampling maps, rather than by explicitly learning a globally invertible coordinate transform.

Here, we implement ACT in a residual and compositional form, inspired by the deformable attention mechanism \citep{zhu2020deformableTrans}. Rather than predicting a completely new coordinate system, each layer estimates an incremental coordinate offset relative to the identity mapping.
Formally, given a latent feature map $h$ defined over $\Omega$, the block predicts a collection of head-specific coordinate offsets $\{\Delta_h^{(m)}\}_{m=1}^M$. These offsets define sampling maps through the residual formulation
\begin{equation}
\label{eq:residual_formulation}
    \phi_h^{(m)}(\xi) = \xi + \Delta_h^{(m)}(\xi), \quad \xi \in \Omega, \quad m = 1, \dots, M,
\end{equation}
where $\Delta_h^{(m)}$ is conditioned on the current feature map. Under this residual parameterization, $\Delta_h^{(m)} \equiv 0$ recovers the identity sampling pattern. Therefore, by controlling the magnitude of the offsets, we can make the realignment easier to control and stabilize.

\begin{figure}
    \centering
    \includegraphics[width=\textwidth]{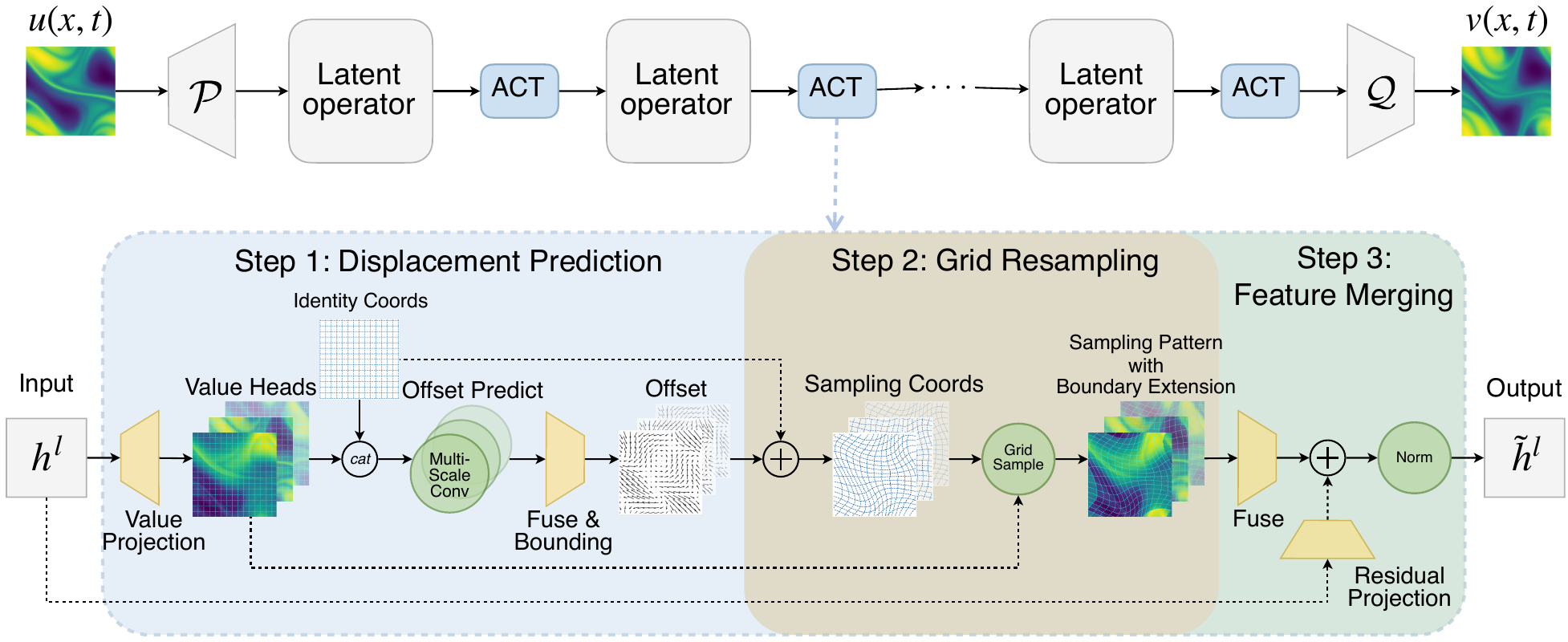}
    \caption{Architecture of the ACT-enhanced operator model.}
    \label{fig:act_block}
\end{figure}

The overall architecture is illustrated in Figure~\ref{fig:act_block}, which can be decomposed into three main steps: coordinate offset prediction, grid resampling, and feature merging.

\textbf{Step 1: Coordinate Offset Prediction.}
We first project the input feature map $h$ and split it into $M$ independent value tensors $\{V_m\}_{m=1}^M$ for head-specific offset prediction.

To incorporate positional information, the normalized base coordinates $\xi$ are concatenated with each value tensor along the channel dimension, forming joint representations $f_{\mathrm{in}}^{(m)} = [V_m, \xi]$.
Instead of using separate predictors for each head, we employ a single \textit{offset predictor} that processes all heads in parallel. The predictor consists of a lightweight multi-scale convolutional module:
\begin{equation}
\Delta_{h, \mathrm{raw}}^{(m)} = \mathrm{Conv}_{\mathrm{fuse}} \Big( \sigma \big( \big[ \mathrm{Conv}_{1\times1}(f_{\mathrm{in}}^{(m)}), \mathrm{Conv}_{3\times3}(f_{\mathrm{in}}^{(m)}), \mathrm{Conv}_{\mathrm{dilated}}(f_{\mathrm{in}}^{(m)}) \big] \big) \Big),
\end{equation}
where $\sigma$ denotes the GELU activation. The predictor outputs a $d$-dimensional offset field that represents coordinate displacements.
To stabilize the realignment, we bounding the raw offsets by a maximum displacement magnitude $\alpha$ as
\begin{equation}
\label{eq::disp}
\Delta_h^{(m)} = \alpha \tanh(\Delta_{h, \mathrm{raw}}^{(m)}).
\end{equation}

\textbf{Step 2: Grid Resampling.}
Using the predicted offsets, we define sampling maps via Eq.~\eqref{eq:residual_formulation}. These maps specify, for each output location $\xi$, the corresponding sampling location $\phi_h^{(m)}(\xi)$ in the original coordinates. The feature representation under the transformed coordinates is obtained by
\begin{equation}
\widetilde{V}_m(\xi) = V_m\big(\phi_h^{(m)}(\xi)\big).
\end{equation}
This operation preserves the underlying feature values while altering the sampling locations, resulting in a different spatial representation of the same feature.

To handle out-of-domain coordinates, we impose physically meaningful boundary conditions. Standard differentiable sampling operators typically handle such coordinates by clamping them to the nearest border. To better respect the underlying physical topology, our ACT block instead maps out-of-bounds sampling locations back into the valid domain prior to the actual grid sampling. For example, in a 1D periodic domain $\Omega = [-1, 1]$, an out-of-bounds sampling coordinate $\phi(\xi)$ is mapped back into the domain via:
\begin{equation}
\hat{\phi}(\xi) = \big( (\phi(\xi) + 1) \bmod 2 \big) - 1.
\end{equation}
This coordinate extension ensures that subsequent sampling queries the physically consistent locations across periodic boundaries, rather than simply replicating edge values. Our experimental results demonstrate that this physics-informed coordinate conditioning prevents boundary artifacts and yields significant performance improvements.

\textbf{Step 3: Feature Merging.}
The sampled features $\widetilde{V}_m$ are concatenated and projected via a pointwise linear layer $W_o$. In parallel, a residual branch preserves the original representation via $h_{\mathrm{res}} = W_{\mathrm{res}} h$. The final output is
\begin{equation}
\widetilde{h} = \mathcal{T}_{\theta}(h) =
\mathrm{Norm}\Big(
h_{\mathrm{res}} +
W_o\big(\mathrm{Concat}(\widetilde{V}_1, \ldots, \widetilde{V}_M)\big)
\Big).
\end{equation}
This residual formulation allows the ACT block to introduce coordinate-level adjustments while retaining the original feature representation, ensuring stable and incremental updates.

\subsection{Instantiation with Neural Operator Backbone}

To prevent spatial misalignment from re-emerging as features propagate through depth, geometric realignment is introduced in a layer-wise manner rather than only once at the input or output. Denoting the operator update at layer $l$ by $\mathcal{G}^{l}_\theta$ and the corresponding adaptive coordinate block by $\mathcal{T}_{\theta}^{l}$, the hidden features evolve as
\begin{equation}
h^{l+1} = \mathcal{T}_{\theta}^{l} \circ \mathcal{G}^{l}_\theta \big(h^{l}\big), \qquad l = 0, \dots, L-1.
\end{equation}
Accordingly, the overall network takes the form
\begin{equation}
\mathcal{F}_\theta = \mathcal{Q} \circ \mathcal{T}_{\theta}^{L-1} \circ \mathcal{G}^{L-1}_\theta \circ \cdots \circ \mathcal{T}_{\theta}^{0} \circ \mathcal{G}^{0}_\theta \circ \mathcal{P}.
\end{equation}

In our implementation, each latent operator block is followed by an ACT block, so the hidden stack alternates between operator propagation and geometric realignment. The lifting and projection operators $\mathcal{P}$ and $\mathcal{Q}$ are standard pointwise layers; further details on resampling under learned coordinates and regularization are deferred to Appendix \ref{app:implementation_details}.

\section{Experiments}

This section first describes the experimental setting, including the datasets, baselines, evaluation metric, and ACT-specific design choices, and then presents the main evaluation and ablation results.

\subsection{Experimental Setting}

\paragraph{Dataset Description:}
To comprehensively evaluate model performance across both two-dimensional and three-dimensional settings, we conduct experiments on six representative datasets drawn from three widely used sources. Specifically, \ns follows the benchmark setting introduced in FNO~\citep{li2021fourier}; \react and \shallow are taken from PDEBench~\citep{pdebench}; and \actmatter, \trltwo, and \MHD(a 3D dataset) are selected from The Well~\citep{thewell}. This diverse suite allows us to assess model robustness across different spatial dimensions and physical regimes under consistent benchmark protocols. In all cases, we aim to learn an operator
$
\mathcal{F}_\theta: \big(u(\cdot,t_0), \ldots, u(\cdot,t_{k})\big) \mapsto u(\cdot, t_{k+1}),
$
which maps a history window of $k+1$ observed states to the following state. 
Detailed dataset statistics are summarized in Table~\ref{tab:dataset_counts}, and full dataset descriptions are provided in Appendix~\ref{app::dataset}.

\paragraph{Baseline Models:}
To rigorously evaluate the potential of our proposed ACT mechanism, we integrate it into three representative state-of-the-art baselines spanning distinct paradigms: CNextU~\citep{convunet} (a cutting-edge convolutional architecture widely adopted in image processing), Transolver++~\citep{wu2024transolverfasttransformersolver,luo2025transolver} (an attention-based operator), and FNO~\citep{li2021fourier} (a classic spectral operator). By applying ACT to these diverse baseline models under identical experimental protocols, we ensure a fair and systematic assessment of its universal effectiveness and modeling enhancements.

\paragraph{Metrics:}
We evaluate model performance using the \emph{Normalized Root Mean Squared Error} (NRMSE), which measures the discrepancy between the predicted field $x$ and the ground-truth field $y$ while being invariant to the scale of the target field. The metric is defined as
\begin{equation}
\mathrm{NRMSE}(x, y)
=
\sqrt{
\frac{\sum_{i=1}^{N}|x_i-y_i|^2}
{\sum_{i=1}^{N}|y_i|^2 + \varepsilon}
},
\end{equation}
where $x_i$ and $y_i$ denote the predicted and ground-truth values at the $i$-th spatial location, respectively, $N$ is the total number of spatial grid points, and $\varepsilon = 10^{-7}$ is a small constant introduced for numerical stability.

\paragraph{Our Design:}
Building upon the baseline models, we augment each backbone with ACT blocks that perform data-driven coordinate adaptation. Each ACT block introduces only a modest number of additional parameters and incurs very limited overhead for large operator backbones such as FNO and CNextU. These modules are interleaved with the backbone layers so that intermediate features are repeatedly represented in adaptive coordinate systems, while preserving the original inductive biases of each baseline architecture. For hierarchical architectures such as CNextU, where the feature-map resolution changes across depth, we use a depth-adaptive maximum displacement magnitude(i.e., $\alpha$ in Eq.~\eqref{eq::disp}), that scales with the local grid spacing. Detailed architectural and training configurations are deferred to Appendix~\ref{app::model}. Finally, to ensure a fair and rigorous evaluation, all other model settings---including network depth, hidden width, training protocols, and optimization hyperparameters---remain strictly identical to those of the corresponding vanilla baselines.

\begin{table}[tbp]
    \centering
    \caption{\textbf{Model Performance Comparison.}
    NRMSE are reported.
    Gray columns correspond to models augmented with ACT.
    Values marked with $\downarrow$ indicate the relative error reduction compared to the vanilla (VAN) setting. 
    The symbols \textperthousand~and \textpertenthousand~denote parts per thousand ($10^{-3}$) and parts per ten thousand ($10^{-4}$), respectively. The best results are highlight in \textbf{Bold}.}
    \resizebox{0.95\columnwidth}{!}{\begin{tabular}{
        @{}l
        c>{\columncolor{gray!20}}c
        c>{\columncolor{gray!20}}c
        c>{\columncolor{gray!20}}c@{}
    }
        \toprule
        \multirow{2}{*}{\texttt{Dataset}}
        & \multicolumn{2}{c}{CNextU}
        & \multicolumn{2}{c}{Transolver}
        & \multicolumn{2}{c}{FNO} \\
        \cmidrule(lr){2-3}
        \cmidrule(lr){4-5}
        \cmidrule(lr){6-7}
        & VAN & +ACT
        & VAN & +ACT
        & VAN & +ACT \\
        \midrule

        \ns
        & 6.13\% & \textbf{4.96\%{\scriptsize$\downarrow$19.1\%}}
        & 32.29\% & \textbf{1.35\%{\scriptsize$\downarrow$95.8\%}}
        & 2.22\% & \textbf{0.62\%{\scriptsize$\downarrow$72.1\%}} \\
        \midrule

        \react
        & 4.03 \textperthousand &  \textbf{1.20 \textperthousand{\scriptsize$\downarrow$70.2\%}}
        & 2.20 \textperthousand & \textbf{0.69 \textperthousand{\scriptsize$\downarrow$68.6\%}}
        & 1.66 \textperthousand & \textbf{1.34 \textperthousand{\scriptsize$\downarrow$19.3\%}} \\
        \midrule

        \shallow
        & 2.11 \textpertenthousand & \textbf{1.58 \textpertenthousand{\scriptsize$\downarrow$25.1\%}  }      & 3.49 \textpertenthousand & \textbf{0.78 \textpertenthousand{\scriptsize$\downarrow$77.7\%}}
        & 1.38 \textpertenthousand & \textbf{0.93 \textpertenthousand{\scriptsize$\downarrow$32.6\%}} \\
        \midrule

        \actmatter
        & 3.43\% & \textbf{2.31\%{\scriptsize$\downarrow$32.7\%}}
        & 18.61\% & \textbf{4.91\%{\scriptsize$\downarrow$73.6\%}}
        & 4.49\% & \textbf{2.69\%{\scriptsize$\downarrow$40.1\%}} \\
        \midrule

        \trltwo
        & 9.14\% & \textbf{7.57\%{\scriptsize$\downarrow$17.2\%}}
        & 20.34\% & \textbf{6.98\%{\scriptsize$\downarrow$65.7\%}}
        & 11.97\% & \textbf{8.95\%{\scriptsize$\downarrow$25.2\%}} \\
        \midrule

        \MHD
        & 12.08\% & \textbf{9.63\%{\scriptsize$\downarrow$20.3\%}}
        & 29.30\% & \textbf{6.82\%{\scriptsize$\downarrow$76.7\%}}
        & 15.40\% & \textbf{9.09\%{\scriptsize$\downarrow$41.0\%}} \\
        \midrule
        
        \rowcolor{white} 
        \textbf{Average $\downarrow$} 
        & --- & \textbf{\scriptsize$\downarrow$30.8\%}
        & --- & \textbf{\scriptsize$\downarrow$76.4\%}
        & --- & \textbf{\scriptsize$\downarrow$38.4\%} \\
        \bottomrule
    \end{tabular}}
    \label{tab:act_column_colored}
\end{table}

\subsection{Evaluation Results}

\paragraph{Benchmark Comparison:}
As shown in Table~\ref{tab:act_column_colored}, we compare the inference performance of three representative backbones (CNextU, Transolver, and FNO), with and without the proposed ACT module, across six PDE datasets. ACT consistently reduces NRMSE across all backbone-dataset pairs in this benchmark. These improvements hold across distinct architectural families, including the convolutional CNextU, the attention-based Transolver, and the spectral FNO. The average relative error reductions reach 76.4\% for Transolver, 38.4\% for FNO, and 30.8\% for CNextU. Overall, these results show that ACT is a broadly effective, model-agnostic module for improving neural operator accuracy across diverse physical systems. For more comprehensive visualization results, please refer to Appendix~\ref{app::add_vis_exp_res}.

\begin{figure}
    \centering
    \includegraphics[width=1.0\linewidth]{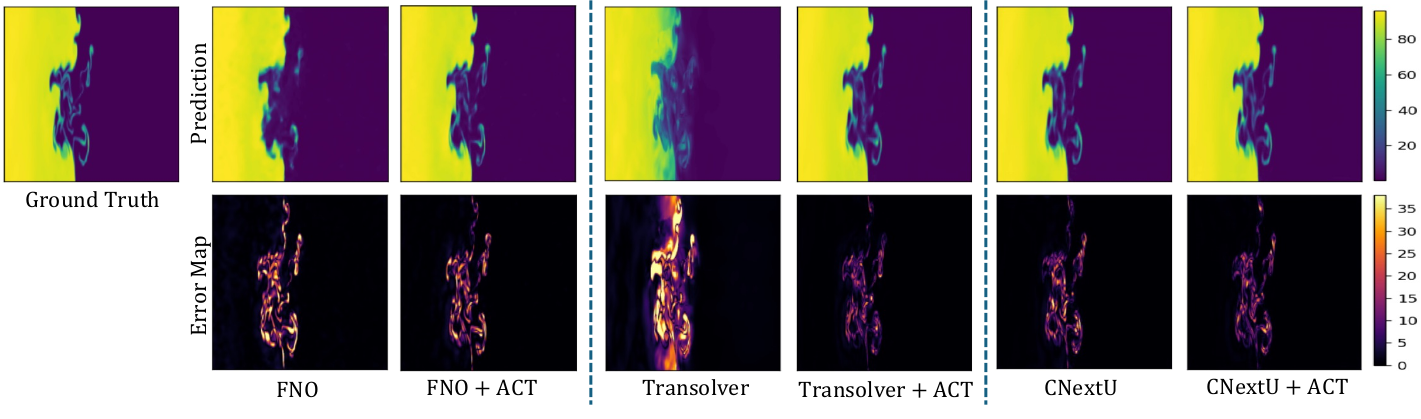}
    \caption{Predicted density $\rho$ results of \trltwo dataset for different models.}
    \label{fig:predictions_turb2d}
\end{figure}





\paragraph{Ablation Study:} We conduct a controlled ablation study on the Navier--Stokes equation using FNO as the backbone to disentangle three factors: increased model capacity, a single terminal realignment, and layer-wise realignment throughout the network. The results are summarized in Table~\ref{tab:ablation}. Specifically, we compare four variants: (i) the vanilla FNO (around 16.8M parameters), (ii) a larger FNO with approximately twice the number of parameters (around 37.9M), (iii) FNO augmented with only one ACT block at the last layer, and (iv) FNO with ACT blocks applied at every layer. This comparison isolates whether the observed gains arise merely from more parameters or from where ACT is inserted.

The results show that simply increasing model capacity yields only limited improvement. In contrast, introducing only a final ACT block already produces a substantially larger gain. Applying ACT blocks in a layer-wise manner yields a further improvement over both the larger-capacity baseline and the final-ACT-only variant. This controlled comparison indicates that the performance gains primarily stem from the introduction of coordinate adaptation, rather than merely increasing model capacity.

\begin{table}
\centering
\caption{Ablation study on the Navier--Stokes equation using FNO as the backbone. Introducing adaptive coordinate transformation (ACT) modules significantly improves performance compared to increasing model capacity.}
\label{tab:ablation}
\resizebox{0.8\columnwidth}{!}{\begin{tabular}{ccccc}
\toprule
Method & FNO & FNO (2$\times$ params) & FNO + Final ACT & FNO + Layer-wise ACT \\
\midrule
NRMSE & 2.22\% & 1.80\% & 1.15\% & \textbf{0.62\%} \\
\bottomrule
\end{tabular}}
\end{table}
\section{Insights}

\begin{figure}
    \centering
    \includegraphics[width=0.7\linewidth]{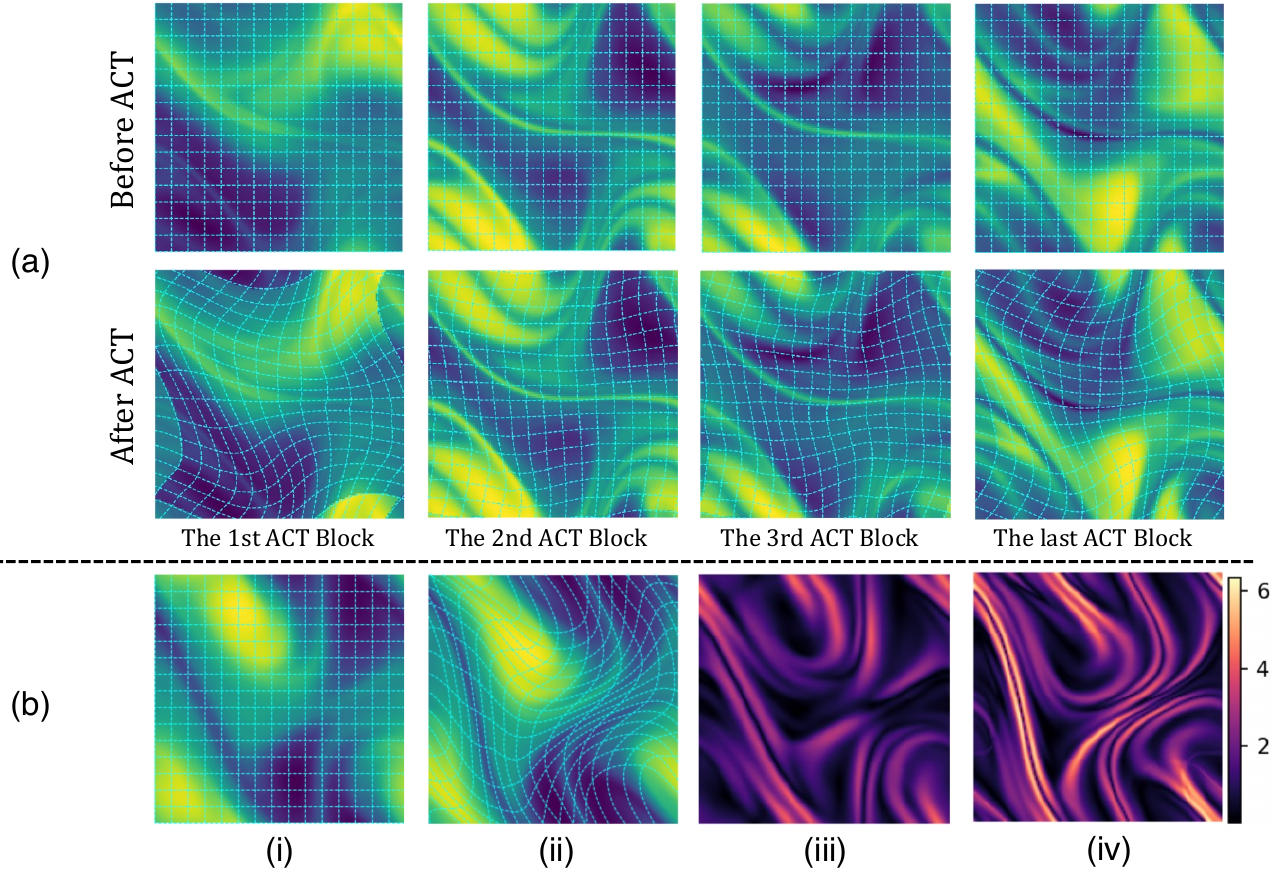}
    \caption{Visualization of feature representations induced by ACTs. (i) Layer-wise visualization of intermediate features, showing that the effect of ACT varies across depth, with both the first and the terminal blocks producing clear realignment. (ii) Features and gradient norms before and after adding only the final ACT block, showing that even this isolated terminal realignment sharpens local structure. Together, the two panels provide qualitative evidence that ACT first simplifies the latent operator mapping and then uses the final block to recover sharp detail.}
    \label{fig:deformation_illustration}
\end{figure}

ACT delivers strong gains across architectures and datasets, which naturally raises a mechanistic question: what exactly are the ACT blocks doing inside the network? To better understand this, we visualize the layer-wise feature evolution and the effect of adding only the terminal ACT block in Figure~\ref{fig:deformation_illustration}. These visualizations suggest that ACT does not play a uniform role across depth. Instead, different blocks appear to divide their labor: earlier blocks mainly reorganize the latent representation, whereas the terminal block is more directly tied to the recovery of sharp local structure. Below, we unpack these two roles in turn.

\subsection{How Intermediate ACT Blocks Simplify the Latent Operator}
\label{sec:operator_structure}

Earlier ACT blocks primarily serve as geometric realignment modules: instead of forcing every latent operator update to model transport or deformation on a fixed grid, they absorb part of this burden into learned coordinate adaptation, so that subsequent operator updates act on hidden representations with more regular spatial structure. In this way, they mitigate spatial misalignment before it accumulates across depth.

To illustrate this effect, we consider a simple setting of an integral operator:
\begin{equation}
\mathcal{F}: u \mapsto v, \qquad
v(x) = \int_\Omega K(x,y)\, u(y)\, dy.
\end{equation}
Next, we examine how input-side and output-side coordinate transformations modify the kernel representation of this operator. Detailed derivations are deferred to Appendix~\ref{app:operator_structure}.

\paragraph{Input-side transformation.}
For an input-side coordinate transformation written as $y = \phi_{\mathrm{in}}(\xi)$, define the transformed input by $\bar{u}(\xi) = u\big(\phi_{\mathrm{in}}(\xi)\big)$. The corresponding effective kernel is
\begin{equation}
\tilde{K}_{\mathrm{in}}(x,\xi) = K\big(x, \phi_{\mathrm{in}}(\xi)\big)\, \left|\det \nabla \phi_{\mathrm{in}}(\xi)\right|,
\end{equation}
Thus, input-side transformations re-parameterize the integration variable and change how information is aggregated from the input domain.

\paragraph{Output-side transformation.}
For an output-side coordinate transformation written as $x = \phi_{\mathrm{out}}(\xi)$, define the transformed output by $\bar{v}(\xi) = v\big(\phi_{\mathrm{out}}(\xi)\big)$. The effective kernel becomes
\begin{equation}
\tilde{K}_{\mathrm{out}}(\xi,y) = K\big(\phi_{\mathrm{out}}(\xi), y\big).
\end{equation}
Hence, the transformation re-parameterizes the evaluation variable and changes where the operator is queried.

\paragraph{Summary.}
Input-side and output-side transformations therefore modify the operator in complementary ways: one changes the aggregation coordinates, while the other changes the query coordinates. In both cases, the original operator $\mathcal{F}$ is rewritten in transformed coordinates as an equivalent operator with a modified kernel representation. This is the mechanism by which geometric realignment can simplify the operator seen by subsequent latent updates. 

\subsection{How the Final ACT Block Restores Sharp Local Structure}

Once earlier ACT blocks have reduced spatial misalignment in the latent features, the final block can focus on restoring sharp local structure near the output. This is consistent with the visualization above that ACT exhibits a depth-dependent division of labor: earlier ACT blocks mainly simplify the latent operator mapping, whereas the last block is important for reconstructing local detail.

Panel (ii) of Figure~\ref{fig:deformation_illustration} provides a more direct check of this idea by isolating the case in which only the final ACT block is added. Even in this restricted setting, the final block visibly sharpens the feature representation and its gradient structure, providing qualitative evidence that terminal realignment contributes directly to recovering sharp local transitions.

One plausible mechanism for this sharpening effect is local coordinate compression. If the final ACT block induces an output-side re-parameterization $\bar{v}(\xi) = v\big(\phi(\xi)\big)$, then the chain rule gives
\begin{equation}
\nabla_{\xi} \bar{v}(\xi) = \nabla \phi(\xi)^\top \, \nabla_x v\big(\phi(\xi)\big).
\end{equation}
Therefore, when $\phi$ locally compresses coordinates along a certain direction, the Jacobian matrix $\nabla \phi(\xi)$ acts as a multiplier on the physical gradient and can increase the magnitude of the corresponding derivative in the transformed representation. This provides a simple explanation for why the terminal ACT block can recover sharper local transitions.

The ablation results in Table~\ref{tab:ablation} further support this interpretation. Adding only a final ACT already improves performance, while layer-wise ACT yields further gains beyond both the vanilla backbone and a larger-capacity baseline. Together, these results suggest that ACT distributes its effect across depth: intermediate blocks reduce geometric complexity, whereas the final block is especially important for recovering local sharpness.

\section{Limitations}
Despite these encouraging results, several aspects of ACT deserve further discussion.
First, both the benchmark results and the analyses suggest that ACT is most beneficial when spatial misalignment, transport, mixing, deformation, or sharp moving interfaces are major sources of difficulty. In such settings, geometric realignment can simplify the latent operator while the terminal ACT block helps recover local sharpness. By contrast, for problems whose dynamics are already well represented in fixed Eulerian coordinates or do not exhibit pronounced geometric variability or sharp transitions, the improvement from ACT may be less significant.

Second, although our visualizations and ablation results support the interpretation that adaptive coordinate transformations simplify the effective operator seen by subsequent latent updates, this claim is currently supported mainly by empirical evidence rather than by a formal theoretical characterization. A more rigorous understanding of when and why learned coordinate realignment improves operator learning remains an important direction for future work.

\section{Conclusion}

In this work, we introduced ACT as a lightweight and plug-and-play mechanism for geometric realignment in neural operator learning. Across multiple backbones and PDE benchmarks, the results show that adaptive coordinate transformations can substantially improve accuracy, especially when transport, mixing, or deformation is a dominant source of complexity. The qualitative analyses and ablations further suggest that ACT helps redistribute part of the geometric burden from the operator updates into the learned coordinate system.

Several directions remain open for future work. In particular, it is important to better understand the theoretical conditions under which learned coordinate realignment simplifies operator learning, and to extend the current framework beyond the present structured-grid setting to more general boundary conditions, complex geometries, and irregular meshes.

More broadly, our results suggest that coordinate choice can fundamentally shape the difficulty of operator learning, and the geometric realignment achieved by ACT offers a promising way to reduce this complexity.

\bibliographystyle{plainnat}
\bibliography{reference}

\newpage
\appendix
\textbf{\large{Appendix}}


\section{Additional Implementation Details} \label{app:implementation_details}

This appendix records implementation details that complement, rather than repeat, the ACT formulation in Section~3.3. In the current implementation, the backbone is instantiated as a 2D Fourier neural operator. The input is lifted by two pointwise convolutions into a hidden representation, processed by a stack of Fourier blocks, and then projected back to the output channels by another pointwise head. Each Fourier block consists of a spectral convolution, a skip connection, optional normalization, and an optional channel-mixing MLP. The implementation also supports gradient checkpointing on the Fourier blocks and an ablation mode in which ACT is disabled by replacing it with a pass-through module.

At the implementation level, each ACT block first applies a $1 \times 1$ projection to produce head-wise value features, then folds the head dimension into the batch dimension so that all heads are processed in parallel. If $x^{(l)}$ denotes the hidden feature map at layer $l$, the packed head features can be written as
\begin{equation}
V^{(l)} = \mathrm{reshape}\big(W_v x^{(l)}\big) \in \mathbb{R}^{(B M) \times C_h \times H \times W}.
\end{equation}
When coordinate conditioning is enabled, the cached identity-grid channels are concatenated to these head-wise value features before offset prediction. In the default branched setting, the displacement tower combines parallel $1 \times 1$, $3 \times 3$, and dilated $3 \times 3$ convolutions before a final $1 \times 1$ fusion layer outputs the offsets. The resulting displacement is bounded by
\begin{equation}
\Delta^{(l)} = \alpha \, \tanh\!\left(f_{\mathrm{disp}}\big([V^{(l)}, X_{\mathrm{id}}]\big)\right),
\end{equation}
where $\alpha$ is the prescribed maximum displacement. The last fusion layer is initialized to zero, so each ACT block starts exactly from the identity transformation. The code also includes a non-branched alternative in which the displacement tower is replaced by a two-layer $3 \times 3$ stack with GroupNorm and GELU.

The reference coordinates are cached as a normalized identity grid on $[-1,1]^2$ and reused across calls as long as the spatial resolution and device do not change. The maximum displacement can either be specified manually or derived from the spatial resolution as the normalized size of one grid cell.

Resampling is carried out with bilinear \texttt{grid\_sample} on normalized coordinates. If $G^{(l)} = X_{\mathrm{id}} + \Delta^{(l)}$ denotes the deformed sampling grid, the code applies a periodic fold only where coordinates are strictly outside the valid range,
\begin{equation}
\widetilde{G}^{(l)} =
\begin{cases}
G^{(l)}, & \text{if } G^{(l)} \in [-1,1]^2, \\
\mathrm{remainder}\big(G^{(l)} + 1, 2\big) - 1, & \text{otherwise.}
\end{cases}
\end{equation}
Hence coordinates already equal to $\pm 1$ remain unchanged, so zero predicted displacement reproduces the exact identity sampling pattern. The actual sampling call uses \texttt{padding\_mode=border} and \texttt{align\_corners=True}.

After resampling, the transformed head features are concatenated and projected back to the hidden width, then merged with a residual shortcut,
\begin{equation}
h^{(l+1)} =
\begin{cases}
\mathrm{Norm}\!\left(W_o\,\widehat{V}^{(l)} + W_{\mathrm{res}} x^{(l)}\right), & \text{for non-final ACT blocks}, \\
W_o\,\widehat{V}^{(l)} + W_{\mathrm{res}} x^{(l)}, & \text{for the final ACT block},
\end{cases}
\end{equation}
where $\widehat{V}^{(l)}$ denotes the concatenated sampled head features. This matches the implementation flag that skips output normalization in the last ACT block. A further training-related detail is that displacement tensors are not returned in training mode: each ACT block outputs \texttt{(tensor, None)} while \texttt{self.training} is true, and only exposes the displacement fields in evaluation mode. This design reduces the need to retain layer-wise displacement tensors during training.

\section{Dataset Ilustrations}
\label{app::dataset}
Table~\ref{tab:dataset_counts} provides a comprehensive overview of the datasets utilized in our experiments. We evaluate our models across a diverse suite of physical systems, encompassing both two-dimensional and three-dimensional Cartesian coordinates. The table details the fundamental properties of each dataset, including the spatial resolution and the temporal length of each trajectory (\texttt{n\_steps}). Furthermore, to ensure rigorous and reproducible benchmarking, we explicitly outline the data partitioning strategy, providing the exact number of trajectories allocated for the training, validation, and test splits.
\begin{table}[h!]
\small
\centering
  \caption{Detailed description of the datasets, including coordinate system (CS), spatial resolution, number of time-steps per trajectory (\texttt{n\_steps}), and the number of trajectories in the training, validation, and test sets.}
 \begin{tabular}{l c r r rrr} 
  \toprule
 \texttt{Dataset} & CS & Resolution & \texttt{n\_steps} & Train & Valid & Test \\
  \midrule
  \small{\ns}        & \cartesiantwoD   & $128\times 128$            & $50$ & $400$ & $50$ & $50$ \\ 
  \small{\react}     & \cartesiantwoD   & $128\times 128$            & $101$ & $800$ & $100$ & $100$ \\ 
  \small{\shallow}   & \cartesiantwoD   & $128\times 128$            & $101$ & $800$ & $100$ & $100$ \\ 
  \small{\actmatter} & \cartesiantwoD   & $256\times 256$            & $81$  & $175$ & $24$  & $26$  \\
  \small{\trltwo}    & \cartesiantwoD & $128\times 384$ & $101$ & $72$  & $9$    & $9$   \\
  \small{\MHD}       & \cartesianthreeD & $64^3$                     & $100$ & $77$  & $10$  & $10$  \\
 \bottomrule
 \end{tabular}
 \vspace{0.5em}
 \label{tab:dataset_counts}
\end{table}

\paragraph{Navier-Stokes Equation: }
Our benchmark includes the two-dimensional Navier-Stokes (NS) equations governing a viscous and incompressible fluid. The dynamics are expressed via the vorticity formulation on a unit torus domain:
\begin{align}
\begin{split}
\partial_t w(x,t) + u(x,t) \cdot \nabla w(x,t) &= \nu \Delta w(x,t) + f(x), \qquad x \in (0,1)^2, t \in (0,T]  \\
\nabla \cdot u(x,t) &= 0, \qquad \qquad \qquad \qquad \quad x \in (0,1)^2, t \in [0,T]  \\
w(x,0) &= w_0(x), \qquad \qquad \qquad \quad x \in (0,1)^2 
\end{split}
\end{align}
In these governing equations, $u \in C([0,T]; H^r_{\text{per}}((0,1)^2; \mathbb{R}^2))$ (with $r>0$) defines the fluid velocity, while $w = \nabla \times u$ denotes the scalar vorticity field. The initial condition is given by $w_0 \in L^2_{\text{per}}((0,1)^2;\mathbb{R})$, $\nu \in \mathbb{R}_+$ controls the kinematic viscosity, and $f \in L^2_{\text{per}}((0,1)^2;\mathbb{R})$ acts as a time-independent external forcing term. We aim to approximate the solution operator $G^\dagger: \omega(\cdot, t_0:t_k) \to \omega(\cdot, t_{k+1})$. 

We evaluate the models under a representative viscosity regime with $\nu = 10^{-4}$. For all experiments, both training and testing data are defined on a fixed spatial grid of resolution $128 \times 128$.

\paragraph{2D Diffusion-Reaction Equation}:
\label{sec:2d_diff-react}
Our study also employs the two-dimensional diffusion-reaction datasets provided by PDEBench~\citep{pdebench}. This benchmark broadens the physical modeling to a 2D spatial domain and involves two non-linearly coupled quantities: the activator $u=u(t,x,y)$ and the inhibitor $v=v(t,x,y)$. The system's dynamics are governed by the following partial differential equations:
\begin{align} \label{eq:2d_diff-react}
    \partial_t u  = D_u \partial_{xx} u + D_u \partial_{yy} u + R_u, 
    ~~~ \partial_t v  = D_v \partial_{xx} v + D_v \partial_{yy} v + R_v \, ,
\end{align}
Here, $D_u$ and $D_v$ represent the constant diffusion coefficients associated with the activator and inhibitor, respectively. Meanwhile, $R_u=R_u(u,v)$ and $R_v=R_v(u,v)$ correspond to their localized reaction terms. The spatio-temporal domain is bounded by $x, y \in (-1,1)$ and spans a time interval of $t \in (0,5]$. Such coupled systems are widely recognized for their utility in simulating biological pattern generation. 
Specifically, the reaction mechanisms are modeled after the classic Fitzhugh-Nagumo kinetics \citep{Klaasen1984fhn}, which are formulated as:
\begin{align}
\label{eq:react_u_diff-sorp}
    R_u(u,v) &= u - u^3 - k - v, \\
\label{eq:react_v_diff-sorp}
    R_v(u,v) &= u - v,
\end{align}
For these experiments, the scalar parameter is fixed at $k = 5 \times 10^{-3}$, and the respective diffusion rates are specified as $D_u = 1 \times 10^{-3}$ and $D_v = 5 \times 10^{-3}$. Initial states are synthesized by sampling from a standard normal distribution, such that $u(0,x,y) \sim \mathcal{N}(0,1.0)$ across the spatial grid. While the native PDEBench repository supplies high-resolution data with grid dimensions of $N_x = 512$, $N_y = 512$, and $N_t = 501$, we adopt the spatially and temporally downsampled variant ($N_x = 128$, $N_y = 128$, $N_t = 101$) to facilitate efficient neural network training. To generate the ground truth, the spatial derivatives were discretized via the finite volume method \citep{Moukalled2016}, while temporal progression was handled by the standard fourth-order Runge-Kutta (RK4) integrator available in the \emph{scipy} library \citep{2020SciPy-NMeth}. Overall, the core objective is to learn the solution operator $G^\dagger$ that maps the coupled historical states of the activator and inhibitor up to an initial time window to their subsequent evolution up to the next timestep. Formally, this operator we aim to learn is defined as $G^\dagger: C([t_0:t_k]; H^r((-1,1)^2; \mathbb{R}^2)) \to C(t_{k+1}; H^r((-1,1)^2;\mathbb{R}^2))$. 


\paragraph{Shallow Water Equations:}
For the shallow-water experiments, we adopt the benchmark configurations provided by PDEBench~\citep{pdebench}. Formulated as a depth-averaged simplification of the compressible Navier-Stokes equations, the shallow-water equations (SWE) serve as a fundamental mathematical model for simulating free-surface fluid dynamics. Mathematically, this 2D hyperbolic PDE system is naturally represented in a vectorized conservation form:
\begin{subequations}
\label{eq:swe}
\begin{align}
    &\partial_t h + \nabla \cdot (h \mathbf{u}) = 0, \\
    &\partial_t (h \mathbf{u}) + \nabla \cdot 
    \left( h \mathbf{u} \otimes \mathbf{u} + \frac{1}{2} g_r h^2 \mathbb{I} \right) 
    = -g_r h \nabla b,
\end{align}
\end{subequations}
In this notation, $\mathbf{u} = (u, v)$ captures the two-dimensional velocity vector, $h$ stands for the localized water column depth, and $b$ denotes the underlying bottom topography (bathymetry). Consequently, the product $h\mathbf{u}$ signifies the momentum vector, while $g_r$ indicates the constant of gravitational acceleration. A critical property of this formulation is its strict preservation of mass and momentum conservation, even in the presence of discontinuous shock waves. This trait makes it highly suitable for generating complex, non-linear physical datasets that mirror real-world scenarios such as flash floods and tsunamis. 
By decomposing the vectors along the spatial axes, the governing dynamics can be explicitly formulated as:
\begin{subequations}
 \begin{align}
     \label{eq:swe1}
      \partial_t h + \partial_x hu + \partial_y hv &= 0\, , \\
      \label{eq:swe2}
      \partial_t hu + \partial_x \left( u^2h + \frac{1}{2} g_r h^2 \right) + \partial_y uvh &= - g_r h \partial_x b \, , \\
      \label{eq:swe3}
      \partial_t hv + \partial_y \left( v^2h + \frac{1}{2} g_r h^2 \right)  + \partial_x uvh &= - g_r h \partial_y b \, ,
\end{align}
\end{subequations}
Regarding the specific experimental setup, the PDEBench data collection models a two-dimensional radial dam-break phenomenon. The simulation operates within a bounded square region $\Omega = [-2.5, 2.5]^2$. The initial fluid depth is structured as a centralized cylindrical column, governed by the following piecewise function:
\begin{equation}
    h(t=0,x,y) = \begin{cases}
    2.0, & \text{for} \; r < \sqrt{x^2 + y^2} \\
    1.0, & \text{for} \; r \geq \sqrt{x^2 + y^2}
    \end{cases}
\end{equation}
Here, the initial column radius $r$ is drawn stochastically from a uniform continuous distribution $\mathcal{U}(0.3, 0.7)$. The reference trajectories for these complex fluid interactions were computed via the robust finite volume schemes implemented within the PyClaw software library~\citep{pyclaw}. Specifically, the primary objective is to approximate the solution operator $G^\dagger$ that maps the historical states of the water depth up to an initial time window to its future dynamics up to the next timestep, the operator we aim to learn is defined as $G^\dagger: C([t_0:t_k]; H^r(\Omega; \mathbb{R})) \to C(t_{k+1}; H^r(\Omega;\mathbb{R}))$, tracking the temporal transition $h|_{\Omega \times [t_0:t_k]} \mapsto h|_{\Omega ,t_{k+1}}$, where $\Omega$ denotes the spatial domain.

\paragraph{Active Matter:}

We incorporate active matter datasets derived from the works of \citep{thewell, maddu2024learning}. Fundamentally, active matter comprises autonomous entities---ranging from synthetic micro-swimmers to biological macromolecules---capable of transducing chemical energy into mechanical motion, consequently exerting local active stresses. Within the bulk, these generated forces propagate via mechanisms like hydrodynamic interactions, structural cross-linking, or direct steric hindrances, resulting in intricately complex spatiotemporal behaviors. The specific configurations adopted in our benchmark model a suspension of active particles within a highly viscous medium. This environment induces orientation-coupled viscosity and pronounced long-range interactions, thereby presenting a formidable challenge for operator learning architectures and data-driven surrogate modeling. 

The foundational data generation simulates $N$ elongated active particles (with length $\ell$ and width $b$, satisfying $\ell/b \gg 1$) dispersed throughout a Stokes flow regime of volume $V$. As the system approaches the macroscopic limit, the suspension dynamics are effectively captured by continuum kinetic frameworks tracking the probability density function $\Psi(\mathbf{x},\mathbf{p}, t)$ \citep{ezhilan2013instabilities,gao2017analytical}. The temporal evolution of $\Psi$, which inherently conserves the total particle count, is dictated by the Smoluchowski equation:
\begin{equation}\label{fokker_planck}
   \frac{\partial \Psi}{\partial t} + \nabla_{\mathbf{x}} \cdot \left( \dot{\mathbf{x}}\Psi \right) + \nabla_{\mathbf{p}} \cdot \left( \dot{\mathbf{p}}\Psi \right) = 0,
\end{equation}
\noindent Here, the variables $\dot{\mathbf{{x}}}$ and $\dot{\mathbf{{p}}}$ represent the translational and rotational fluxes, respectively, dictated by individual particle kinematics within the ambient velocity field $\mathbf{u}(\mathbf{x},t)$. Integrating the distribution function over the orientational space ($\langle f \rangle = \int_{|\mathbf{p}|=1} f \Psi ~ d\mathbf{p}$) allows us to extract macroscopic moments: the scalar concentration $c = \langle 1 \rangle$, the polarity vector $\mathbf{n} = \langle\mathbf{p}\rangle/c$, and the nematic order tensor $\mathbf{Q} = \langle \mathbf{p}\mathbf{p}\rangle/c$. In the context of dense particle suspensions, these phase-space fluxes are formulated as:
\begin{align}
 \dot{\mathbf{{x}}} = \mathbf{u} - d_T \nabla_\mathbf{x} \log \Psi; \: \quad \dot{\mathbf{{p}}} = (\mathbf{I} - \mathbf{pp}) \cdot \left( \nabla \mathbf{u} + 2 \zeta \mathbf{D}\right) \cdot \mathbf{p} -d_R \nabla_\mathbf{p} \log \Psi.
\end{align}
In these expressions, $d_T$ and $d_R$ denote the non-dimensional coefficients for translational and rotational diffusion, respectively. The parameter $\zeta$ quantifies the nematic alignment intensity driven by steric repulsion, while $\mathbf{D} = \langle\mathbf{p}\mathbf{p}\rangle$ defines the second-order orientational moment. Furthermore, the fluid hydrodynamics are intricately coupled to the Smoluchowski dynamics via the incompressible Stokes equations:
\begin{align}
\quad & -\Delta \mathbf{u} + \nabla P = \nabla \cdot \mathbf{\Sigma}, \:\:\: \nabla \cdot \mathbf{u} = 0 \label{eq:stokes},  \\
\color{black}\mathbf{\Sigma} &= \alpha \mathbf{D} +  \beta \mathbf{S\!:\!E} \color{black} - 2 \zeta \beta \left(\mathbf{D} \cdot \mathbf{D} - \mathbf{S\!:\!D}\right). \label{eq:stress_tensor}
\end{align}
Within this fluid coupling, $P(\mathbf{x},t)$ stands for the hydrodynamic pressure. The parameters $\alpha$ and $\beta$ scale the non-dimensional active dipole magnitude and the effective particle concentration, respectively. Additionally, $\mathbf{E} = [\nabla \mathbf{u} + \nabla \mathbf{u}^{\top}] /2$ calculates the symmetric strain-rate tensor, and $\mathbf{S} = \langle\mathbf{p}\mathbf{p}\mathbf{p}\mathbf{p}\rangle$ corresponds to the fourth-order orientational moment.

The coupled governing equations~(\ref{fokker_planck})-(\ref{eq:stokes}) were numerically resolved via a pseudo-spectral framework utilizing $256 \times 256$ spatial modes on a doubly periodic domain of size $(0,10)^2$, integrated temporally with a second-order semi-implicit backward differentiation scheme (SBDF2). Key dimensionless parameters were systematically modulated: the active contractility coefficient was uniformly sampled from $\alpha\in\{-1,-2,-3,-4,-5\}$, the effective concentration parameter was fixed at $\beta= 0.8$, and the nematic alignment factor traversed $\zeta\in\{1,3,5,7,9,11,13,15,17\}$. The final curated dataset encapsulates an extensive suite of macroscopic physical fields: scalar concentration, 2D fluid velocity, orientational order tensor, and the symmetric strain-rate tensor.  
Formally, the target operator is defined as $G^\dagger: C([t_0:t_k]; H^r_{\text{per}}((0,1)^2; \mathbb{R}^d)) \to C(t_{k+1}; H^r_{\text{per}}((0,1)^2;\mathbb{R}^d))$, tracking the temporal transition of the combined macroscopic state vector $\boldsymbol{\phi}$: $\boldsymbol{\phi}|_{(0,1)^2 \times [t_0:t_{k}]} \mapsto \boldsymbol{\phi}|_{(0,1)^2 , t_{k+1}}$.


\paragraph{Turbulent Radiative Layer (2D):}

The benchmark additionally incorporates the 2D turbulent mixing layer dataset from The Well~\citep{thewell}, based on the foundational simulations by \citep{Fielding:2020}. This dataset models the interfacial dynamics between cold, dense gas and ambient hot plasma, where the survival of cold structures is dictated by the competition between turbulent mixing and rapid radiative cooling~\citep{Gronke:2018, Abruzzo:2024}. The primary objective is to learn the forward solution operator $G^\dagger: C([t_0:t_{k}]; H^r(\Omega; \mathbb{R}^4)) \to C(t_{k+1}; H^r(\Omega;\mathbb{R}^4))$, which tracks the temporal transition of the coupled fluid fields across the 2D spatial domain $\Omega$: $(\rho, P, \vec{v})|_{\Omega \times [t_0:t_{k}]} \mapsto (\rho, P, \vec{v})|_{\Omega , t_{k+1}}$.

Mathematically, the system governs the hydrodynamic conservation laws augmented with a temperature-dependent energy sink, parameterized by a characteristic cooling time $t_{\rm cool}$:
\begin{align}
\frac{ \partial \rho}{\partial t} + \nabla \cdot \left( \rho \vec{v} \right) &= 0 \\
\frac{ \partial \rho \vec{v} }{\partial t} + \nabla \cdot \left( \rho \vec{v}\vec{v} + P \right) &= 0 \\
\frac{ \partial E }{\partial t} + \nabla \cdot \left( (E + P) \vec{v} \right) &= -\frac{E}{t_{\rm cool}} \\
E &= P / (\gamma -1) \quad \text{where} \quad \gamma = 5/3
\end{align}
A fundamental property of this system is the scaling relationship between the volume-integrated cooling rate, the net mass accretion flux, the relative velocity, and the cooling timescale, expressed as $\dot{E}_{\rm cool} \propto \dot{M} \propto v_{\rm rel}^{3/4} t_{\rm cool}^{-1/4}$. 

To systematically sample this dynamic landscape, the underlying high-fidelity simulations were generated by sweeping the cooling timescale across $$t_{\rm cool} \in \{0.03, 0.06, 0.1, 0.18, 0.32, 0.56, 1.00, 1.78, 3.16\}.$$ The finalized dataset captures the temporal evolution of the 2D density, pressure, and velocity fields, efficiently circumventing the substantial computational overhead required for conventional numerical integration.

\paragraph{Magnetohydrodynamic (MHD) Turbulence:}
We integrate the magnetohydrodynamic (MHD) turbulence dataset directly from The Well repository \citep{thewell}. Turbulence within MHD systems acts as a cornerstone for numerous astrophysical phenomena, dictating processes from interstellar medium (ISM) evolution to large-scale galactic formation. Specifically, this dataset features isothermal, non-self-gravitating MHD flows that accurately emulate the physical environment of the diffuse ISM. The primary objective is to approximate the solution operator $G^\dagger$ that maps the temporal history of the coupled magnetohydrodynamic fields---comprising the scalar density, the 3D fluid velocity vector, and the 3D magnetic field vector---up to an initial time window  to their subsequent turbulent evolution up to a final time. 

The governing ideal MHD equations, resolved via a third-order hybrid essentially non-oscillatory (ENO) scheme \citep{Cho2003}, formulate the conservation laws for mass, momentum, and magnetic induction as follows:
\begin{align}
 \frac{\partial \rho}{\partial t} + \nabla \cdot (\rho \pmb{\varv}) &= 0, \\
 \frac{\partial \rho \pmb{\varv}}{\partial t} + \nabla \cdot \left[ \rho \pmb{\varv} \pmb{\varv} + \left( p + \frac{B^2}{8 \pi} \right) {\bf I} - \frac{1}{4 \pi}{\bf B}{\bf B} \right] &= {\bf f},  \\
 \frac{\partial {\bf B}}{\partial t} - \nabla \times (\pmb{\varv} \times{\bf B}) &= 0. \ 
\end{align}
In this system, $\rho$ signifies the fluid density, $\pmb{\varv}$ defines the local velocity vector, ${\bf B}$ represents the magnetic field, $p$ denotes the isotropic gas pressure ($p = c_{\rm s}^2 \rho$ with isothermal sound speed $c_{\rm s}$), and ${\bf I}$ acts as the identity tensor. Continuous turbulence is sustained by the external source term ${\bf f}$, which operates as a stochastic solenoidal driving force peaking at the wavenumber $k \approx 2.5$. 

To ensure a robust exploration of the chaotic parameter space, the dataset provides a diverse spectrum of turbulent flows systematically modulated across two key dimensionless parameters: the sonic Mach number ($\mathcal{M}_{\rm s} \equiv |\pmb{\varv}|/c_{\rm s} \in\{0.5, 0.7, 1.5, 2.0, 7.0\}$) and the Alfvénic Mach number ($\mathcal{M}_{\rm A} \equiv |\pmb{\varv}|/ \langle \varv_A \rangle \in \{0.7, 2.0\}$). These configurations encompass both magnetically-dominated sub-Alfvénic regimes and super-Alfvénic environments. The generated snapshots, curated at a $64^3$ spatial resolution to balance fidelity with computational tractability, provide a highly comprehensive multi-modal dataset consisting of scalar mass density, vector fluid velocity, and vector magnetic fields. Formally, this operator is defined as $G^\dagger: C([t_0:t_{k}]; H^r_{\text{per}}(\Omega; \mathbb{R}^7)) \to C(t_{k+1}; H^r_{\text{per}}(\Omega;\mathbb{R}^7))$, where $\Omega$ denotes the 3D periodic spatial domain, tracking the temporal transition $(\rho, \pmb{\varv}, {\bf B})|_{\Omega \times [t_0,t_k]} \mapsto (\rho, \pmb{\varv}, {\bf B})|_{\Omega, t_{k+1}}$, where $\Omega$ denotes the 3D periodic spatial domain.

\section{Model Structure and Hyperparameters}
\label{app::model}

\subsection{ACT-CNextU}
\label{app::modelconv}
As detailed in Table~\ref{tab:convnext_detailed_optional}, the proposed Adaptive Coordinate Transformation (ACT) enhances the ConvNeXt-UNet\citep{convunet} architecture by systematically integrating a \texttt{MultiHeadACTBlock} after each encoder and decoder stage. The vanilla ConvNeXt-UNet serves as a robust multi-scale backbone, comprising 18,571,562 trainable parameters. Enabling the ACT mechanism across all 8 hierarchical stages introduces an additional 997,878 parameters. This corresponds to an overhead of approximately 5.37\% relative to the vanilla model, yielding a total of 19,569,440 parameters for the enhanced \texttt{UNetConvNextACT}. This marginal increase confirms that the performance improvements achieved by our method are fundamentally driven by the structured, coordinate-aware spatial transformations introduced by ACT, rather than a mere scaling of the model capacity.

We further emphasize that the integration of ACT is seamlessly adapted to the hierarchical nature of the U-Net architecture. Throughout all experimental evaluations, the underlying multi-scale encoder-bottleneck-decoder structure remains strictly fixed. While the offset generation branches within the ACT modules share a highly efficient, constant parameter footprint across stages, the linear projection layers naturally scale with the corresponding feature map dimensions and channel counts. Consequently, the computational cost introduced by the ACT modules is optimally distributed across the network's resolutions. This demonstrates that ACT is a highly efficient, plug-and-play enhancement that can be gracefully applied to modern hierarchical vision backbones without disrupting their original design principles.

\begin{table}[t]
\centering
\small
\setlength{\tabcolsep}{5pt}
\caption{Detailed parameter breakdown for the ConvNeXt-UNet enhanced with the ACT module (ACT-CNextU). The ACT mechanism is composed of \texttt{MultiHeadACTBlock}s distributed across the encoder and decoder stages. Disabling them recovers the vanilla ConvNeXt-UNet.}
\begin{tabular}{p{0.60\linewidth} p{0.10\linewidth} p{0.20\linewidth}}
\hline
\textbf{Component} & \textbf{Params} & \textbf{Notes} \\
\hline
\multicolumn{3}{l}{\textit{Input / Stem}} \\
Initial convolutions: Conv2d $\times$ 2 & 3,824 & 3,066 + 758 \\
\hline
\multicolumn{3}{l}{\textit{Vanilla ConvNeXt-UNet Architecture (Subtotals)}} \\
Encoder (4 stages): Downsample + 2$\times$ConvNeXt blocks & 3,674,156 & feature encoding \\
Bottleneck: 1$\times$ConvNeXt block & 3,651,648 & latent representation \\
Decoder (4 stages): Conv2d + Upsample + 2$\times$ConvNeXt blocks & 11,241,934 & feature decoding \\
\quad \textbf{Vanilla Backbone subtotal} & \textbf{18,567,738} & multi-scale UNet \\
\hline
\multicolumn{2}{l}{\textit{MultiHeadACTBlock (ACT, distributed across 8 stages)}} & apply ACT \\
\quad value\_proj Conv2d & 485,888 & channel mixing \\
\quad disp branches (ModuleDict): 1x1, 3x3, dilated 3x3 & 22,080 & $8\!\times\!(152\!+\!1304\!+\!1304)$ \\
\quad disp\_fuse: GELU + Conv2d($24\!\to\!2$) & 400 & $8\!\times\!50$ \\
\quad sampled with augmented grid  & --   & \\
\quad out\_proj Conv2d & 485,730 & project features to output \\
\quad GroupNorm & 3,780 & output normalization \\
\quad \textbf{ACT modules subtotal} & \textbf{997,878} & 8 ACT blocks in total \\
\hline
\textbf{Total trainable params (apply ACT)} & \textbf{19,569,440} & UNetConvNextACT \\
\textbf{Total trainable params (vanilla UNet)} & \textbf{18,571,562} & \texttt{use\_ACT=False} \\
\textbf{Optional ACT overhead} & \textbf{+997,878} & \textbf{+5.37\%} \\
\hline
\end{tabular}

\label{tab:convnext_detailed_optional}
\end{table}

\subsection{ACT-Translover}
\label{app::modeltrans}
As detailed in Table~\ref{tab:transolver_detailed_optional}, the proposed Adaptive Coordinate Transformation (ACT) augments the Transolver backbone by integrating a \texttt{MultiHeadACTBlock} after each standard \texttt{TransolverBlock}. Because the vanilla Transolver is an inherently lightweight architecture, comprising only 188,198 trainable parameters, enabling the ACT mechanism introduces an additional 144,104 parameters. Although this corresponds to a 76.57\% relative parameter overhead, the absolute increase remains highly tractable, resulting in a total of just 332,302 parameters for the enhanced model (TransolverPlus). This demonstrates that the performance improvements achieved by our method are fundamentally driven by the structured, coordinate-aware feature transformations introduced by ACT, rather than relying on massive parameter scaling or heavy model capacities.

We further emphasize that, throughout all experimental evaluations, our Transolver-based models strictly adhere to the unified architectural configuration presented in Table~\ref{tab:transolver_detailed_optional}. When deploying the model across different datasets, the core backbone architecture remains completely fixed. The only source of variation in the overall parameter count arises from the input lifting stem and the output projection head, whose dimensions depend solely on the dataset-specific input and output channels. Consequently, the parameter overhead introduced by the ACT blocks is essentially invariant across different datasets. Therefore, reporting a single representative parameter breakdown is sufficient to accurately characterize the architectural modifications and the additional computational cost introduced by ACT under all experimental settings.

\begin{table}[t]
\centering
\small
\setlength{\tabcolsep}{5pt}
\caption{Detailed parameter breakdown for Transolver enhanced with the ACT module(ACT-Transolver). \texttt{MultiHeadACTBlock} acts as the ACT mechanism; disabling it recovers the vanilla Transolver architecture.}
\begin{tabular}{p{0.60\linewidth} p{0.10\linewidth} p{0.20\linewidth}}
\hline
\textbf{Component} & \textbf{Params} & \textbf{Notes} \\
\hline
\multicolumn{3}{l}{\textit{Input / Stem}} \\
Global Parameter (Transolver root) & 64 & global token / state \\
TransolverMLP: Linear + GELU + Linear & 9,664 & 1,408 + 8,256 (feature lifting) \\
\hline
\multicolumn{3}{l}{\textit{Backbone stage (repeated 4 times)}} \\
\textbf{Step $k$ ($k=1\ldots4$):} & -- &  \\
\hline
\multicolumn{3}{l}{\textit{TransolverBlock (single block details)}} \\
\quad LayerNorm & 128 & pre-normalization \\
\quad PhysicsAttention1DEidetic & 9,129 & physics-attention \\
\quad LayerNorm & 128 & pre-normalization \\
\quad TransolverMLP: Linear + GELU + Linear & 33,088 & 16,640 + 16,448 \\
\quad \textbf{TransolverBlock subtotal} & \textbf{42,473} & per block \\
\hline
\multicolumn{2}{l}{\textit{MultiHeadACTBlock (when apply ACT)}} & apply ACT, if disabled, this becomes vanilla Transolver \\
\quad value\_proj Conv2d & 16,640 & $256\!\to\! 64$ \\
\quad disp branches (ModuleDict): 1x1, 3x3, dilated 3x3 & 2,760 & channel mixing \\
\quad disp\_fuse: GELU + Conv2d($24\!\to\!2$) & 50 & offset prediction \\
\quad sampled with augmented grid  & --   \\
\quad out\_proj Conv2d & 16,448 & project features to output channels \\
\quad GroupNorm & 128 & output normalization \\
\quad \textbf{ACTBlock subtotal} & \textbf{36,026} & per block \\
\hline
Backbone total (with ACT): $4\times$(TransolverBlock + ACTBlock) & 313,996 & $4 \!\times \!(42,473\!+\!36,026)$ \\
Backbone total (vanilla Transolver): $4\times$TransolverBlock & 169,892 & ACT disabled \\
\hline
\multicolumn{3}{l}{\textit{Output Head for ACT-Transolver}} \\
Projection: Conv2d + GELU + Conv2d & 8,578 & 8,320 + 258 \\
\multicolumn{3}{l}{\textit{Output Head for Vanilla Transolver}} \\
Projection: LayerNorm + Linear & 258 &  \\
\hline
\textbf{Total trainable params (apply ACT)} & \textbf{332,302} & TransolverPlus \\
\textbf{Total trainable params (vanilla Transolver)} & \textbf{179,878} & \texttt{use\_ACT=False} \\
\textbf{Optional ACT overhead} & \textbf{+144,104} & \textbf{+76.57\%} \\
\hline
\end{tabular}
\label{tab:transolver_detailed_optional}
\end{table}

\subsection{ACT-FNO}
\label{app::modelfno}
\begin{table}[t]
\centering
\small
\setlength{\tabcolsep}{5pt}
\caption{Layer-by-layer parameter breakdown of the proposed ACT-FNO architecture. The \texttt{MultiHeadACTBlock} acts as a lightweight, plug-and-play enhancement; removing it completely recovers the standard vanilla FNO baseline.}
\begin{tabular}{p{0.60\linewidth} p{0.10\linewidth} p{0.20\linewidth}}
\hline
\textbf{Component} & \textbf{Params} & \textbf{Notes} \\
\hline
\multicolumn{3}{l}{\textit{Input / Stem}} \\
Lifting: Conv1x1($C_{in}\!\to\!128$) + GELU + Conv1x1($128\!\to\!64$) & 9,408 & feature lifting \\
\hline
\multicolumn{3}{l}{\textit{Backbone stage (repeated 4 times, alternating)}} \\
\textbf{Step $k$ ($k=1\ldots4$):} & -- &  \\
\hline
\multicolumn{3}{l}{\textit{FNOBlock2d (single block details)}} \\
\quad SpectralConv2d(modes 32, channel 64) & 4,194,304 & dominant spectral weights \\
\quad Conv2d (skip / linear) & 4,096 & channel mixing \\
\quad GroupNorm $\times 2$ & 256 & 128 + 128 \\
\quad MLP: Conv1x1($64\!\to\!32$) + GELU + Conv1x1($32\!\to\!64$) & 4,192 & 2,080 + 2,112 \\
\quad SoftGating & 64 & per-channel gate \\
\quad \textbf{FNOBlock2d subtotal} & \textbf{4,202,912} & per block \\
\hline
\multicolumn{2}{l}{\textit{MultiHeadACTBlock (when apply ACT)}} & apply ACT, if disabled, this becomes vanilla FNO \\
\quad value\_proj Conv1x1 & 4,160 & $64\!\to\!64$ \\
\quad disp branches (ModuleDict): 1x1, 3x3, dilated 3x3 & 2,760 & channel mixing \\
\quad disp\_fuse: GELU + Conv1x1($24\!\to\!2$) & 50 & offset prediction \\
\quad sampled with arguemented grid  & --    \\
\quad out\_proj Conv1x1 & 4,160 & project features to output channels \\
\quad GroupNorm & 128 & output normalization \\
\quad \textbf{ACTBlock subtotal} & \textbf{11,258} & per block \\
\hline
Backbone total (with ACT): $4\times$(FNOBlock2d + ACTBlock) & 16,856,680 & $4\times(4,202,912+11,258)$ \\
Backbone total (vanilla FNO): $4\times$FNOBlock2d & 16,811,648 & ACT disabled \\
\hline
\multicolumn{3}{l}{\textit{Head}} \\
Projection: Conv1x1($64\!\to\!128$) + GELU + Conv1x1($128\!\to\!C_{out}$) & 8,578 & 8,320 + 258 \\
\hline
\textbf{Total trainable params (apply ACT)} & \textbf{16,874,666} & ACT-FNO \\
\textbf{Total trainable params (vanilla FNO)} & \textbf{16,829,634} & \texttt{use\_ACT=False} \\
\textbf{Optional ACT overhead} & \textbf{+45,032} & \textbf{+0.27\%} only \\
\hline
\end{tabular}
\label{tab:deformerfno_detailed_optional}
\end{table}

As summarized in Table~\ref{tab:deformerfno_detailed_optional}, the proposed Adaptive Coordinate Transformation (ACT) augments the FNO backbone by inserting a lightweight \texttt{MultiHeadACTBlock} after each backbone stage, incurring only a marginal increase in trainable parameters. Concretely, enabling the ACT block introduces an additional 45,032 parameters, which corresponds to approximately 0.27\% overhead relative to the vanilla FNO backbone. This clearly indicates that the performance improvements achieved by ACT are not attributed to increased model capacity, but rather to more effective and structured coordinate-aware feature transformations.

We further emphasize that, throughout all experimental evaluations, our FNO-based models strictly adhere to the unified architectural configuration presented in Table~\ref{tab:deformerfno_detailed_optional}. When deploying the model across different datasets, the backbone architecture remains fixed, and the only source of variation in parameter count arises from the input lifting layer and the output projection head, whose dimensions depend solely on the dataset-specific input and output channels. Consequently, the relative parameter overhead introduced by the ACT block is essentially invariant across different datasets. Therefore, reporting a single representative parameter breakdown is sufficient to accurately characterize the additional computational cost introduced by ACT under all experimental settings.

\subsection{Common Experiment Settings}
\label{app::expcommon}
To ensure a rigorous and fair evaluation, all compared models are trained and tested under a strictly unified experimental pipeline. The comprehensive list of common hyperparameters and configuration details is summarized in Table~\ref{tab:common_experiment_settings}. Specifically, the data formulation is fixed to utilize 4 historical steps to predict the subsequent 1 step, with consistent Z-score normalization and temporal strides. Furthermore, the optimization dynamics---including the AdamW optimizer, learning rate schedules, batch size, and total training epochs---are held strictly constant across all baselines and our proposed models. Crucially, by keeping these training and validation protocols identical, we isolate the network architecture as the sole independent variable. This guarantees that any observed performance improvements are strictly attributable to the architectural enhancements (such as the integration of the ACT module) rather than dataset-specific hyperparameter tuning or training tricks.
\begin{table}[t]
\centering
\small
\setlength{\tabcolsep}{8pt}
\caption{Common experimental settings shared by all compared models. Only the model architecture is changed across experiments, ensuring a fair comparison.}
\begin{tabular}{lll}
\hline
\textbf{Category} & \textbf{Hyperparameter} & \textbf{Value} \\
\hline
\multirow{2}{*}{ACT Block}
& maximum displacement magnitude(Transolver\&FNO) & 0.5\\
& maximum displacement magnitude(Unet) & adaptive gird width($\frac{2}{\text{Resolution of Input}}$)\\
\hline
\multirow{4}{*}{Data}
& Input steps & 4 \\
& Output steps & 1 \\
& Normalization & Z-score normalization \\
& Temporal stride & stride = 1 \\
\hline
\multirow{1}{*}{Batching}
& Batch size & 32 \\
\hline
\multirow{3}{*}{Optimization}
& Optimizer & AdamW \\
& Learning rate & $5\times10^{-4}$ \\
& Weight decay & $1\times10^{-4}$ \\
\hline
\multirow{3}{*}{LR schedule}
& Scheduler & StepLR \\
& Step size / decay factor for 2d data & 50 / 0.5 \\
& Step size / decay factor for 3d data & 25 / 0.5 \\
\hline
\multirow{7}{*}{Training}
& Epochs for 2d data except \actmatter & 300 \\
& Epochs for \actmatter & 200 \\
& Epochs 3d data & 100 \\
& Loss function & MSE \\
& Gradient clipping & 1.0 \\
& Formatter & channels-first \\
& Checkpoint frequency & every 10 epochs \\
\hline
\multirow{2}{*}{Validation}
& Validation frequency & every 20 epochs \\
& Short validation length & 30 batches \\
\hline
\end{tabular}
\label{tab:common_experiment_settings}
\end{table}



\subsection{Computing Configuration}
\label{app::computingbg}

As shown in table \ref{tab:hardware_setup}, all models and physical simulations were executed on a high-performance computing (HPC) cluster running a Linux-based operating system. 
The compute nodes are equipped with modern multi-core server-grade processors and large-memory configurations to support data-intensive workloads.

To accelerate neural network training and large-scale data generation, GPU-enabled nodes were utilized. 
Each node is equipped with four NVIDIA A100 Tensor Core GPUs, each providing 80\,GB or 40\, GB of VRAM, enabling efficient distributed and memory-intensive computations.

\begin{table}[h]
    \centering
    \caption{Hardware configuration of the GPU-enabled compute nodes used for model training.}
    \label{tab:hardware_setup}
    \begin{tabular}{ll}
        \toprule
        \textbf{Component} & \textbf{Specification} \\
        \midrule
        Operating System & Linux-based HPC environment \\
        CPU & Multi-core server-grade processors (node-dependent) \\
        GPU & $4 \times$ NVIDIA A100 (80\,GB VRAM or 40\,GB VRAM)  \\
        System Memory & High-memory configuration (node-dependent) \\
        \bottomrule
    \end{tabular}
\end{table}

To evaluate the computational efficiency of our proposed method, Table \ref{tab:gpu_profile_shallow_reaction_bs32} profiles the model parameters, GPU memory footprint, and single-forward inference latency across different baseline architectures on the shallow water dataset. The results demonstrate that the integration of the ACT module introduces merely a marginal increase in learnable parameters (e.g., adding only 0.04M parameters to FNO). While the dynamic coordinate reparameterization inherently incurs an expected overhead in memory consumption and forward inference time, the overall computational cost remains well within a practical range. Given the substantial improvements in predictive accuracy, this modest computational trade-off is highly justified.

\begin{table}[h]
    \centering
    \caption{GPU memory footprint and single-forward inference time on shallow water dataset. All experiments use a batch size of 32. Forward time is the median over 3 measured batches after 1 warmup batch.}
    \label{tab:gpu_profile_shallow_reaction_bs32}
    {
    \begin{tabular}{lrrr}
        \toprule
        \textbf{Model} & \textbf{Params} & \textbf{Memory} & \textbf{Forward} \\
         & \textbf{(M)} & \textbf{(MiB)} & \textbf{(s)} \\
        \midrule
        CNextU & 18.57 & 1120 & 0.038 \\
        CNextU + ACT & 19.57 & 4278 & 0.072 \\
        FNO & 16.83 & 1410 & 0.029 \\
        FNO + ACT & 16.87 & 1812 & 0.054 \\
        Transolver & 0.18 & 3640 & 0.128 \\
        Transolver + ACT & 0.33 & 4606 & 0.210 \\
        \bottomrule
    \end{tabular}
    }
\end{table}
\section{Derivation of Transformed Operator Representations}
\label{app:operator_structure}

This appendix expands the discussion in Section~\ref{sec:operator_structure} and shows explicitly how coordinate transformations modify the kernel representation of an integral operator. The notation follows Section~\ref{sec:method}: $\psi_{\mathrm{in}}$ and $\psi_{\mathrm{out}}$ denote forward coordinate transformations, while $\phi_{\mathrm{in}} := \psi_{\mathrm{in}}^{-1}$ and $\phi_{\mathrm{out}} := \psi_{\mathrm{out}}^{-1}$ denote the associated inverse maps. In the ACT instantiation of Section~3.3, these abstract transformations are realized implicitly through the layer-wise blocks $\mathcal{T}^{(l)}_\theta$ and their head-specific sampling maps $\phi_h^{(m)}$.

Consider
\begin{equation}
\mathcal{F}: u \mapsto v, \qquad
v(x) = \int_\Omega K(x,y)\, u(y)\, dy.
\end{equation}
Throughout this appendix, we assume that the coordinate transformations are bijective and sufficiently smooth so that the standard change-of-variables formula applies. The goal is not to change the underlying operator itself, but to rewrite the same mapping in different coordinates and make explicit how the kernel representation is modified.

\paragraph{Input-side transformation.}
Let $\psi_{\mathrm{in}}: \Omega \to \Omega$ be an input-side coordinate transformation, define transformed coordinates by $\xi = \psi_{\mathrm{in}}(y)$, and let $\phi_{\mathrm{in}} := \psi_{\mathrm{in}}^{-1}$ denote the associated inverse map. The transformed input is
\begin{equation}
\bar{u}(\xi) = u\big(\phi_{\mathrm{in}}(\xi)\big).
\end{equation}
Equivalently, $u(y) = \bar{u}(\psi_{\mathrm{in}}(y))$, so the operator can be viewed as acting on the pulled-back input $\bar{u}$ instead of the original field $u$. Since only the source variable $y$ is re-parameterized, the evaluation coordinate $x$ remains unchanged. Applying the change of variables $y = \phi_{\mathrm{in}}(\xi)$, which leads to 
\begin{equation}
dy = \left|\det \nabla \phi_{\mathrm{in}}(\xi)\right| d\xi,
\end{equation}
where $\nabla \phi_{\mathrm{in}}$ is the Jacobian matrix of $\phi_{\mathrm{in}}$, and $\det$ denotes the determinant. Substituting this relation into the operator gives
\begin{equation}
\begin{aligned}
v(x)
&= \int_\Omega K(x,y)\, u(y)\, dy \\
&= \int_\Omega K\big(x, \phi_{\mathrm{in}}(\xi)\big)\, u\big(\phi_{\mathrm{in}}(\xi)\big)\, \left|\det \nabla \phi_{\mathrm{in}}(\xi)\right|\, d\xi \\
&= \int_\Omega K\big(x, \phi_{\mathrm{in}}(\xi)\big)\, \bar{u}(\xi)\, \left|\det \nabla \phi_{\mathrm{in}}(\xi)\right|\, d\xi \\
&= \int_\Omega \tilde{K}_{\mathrm{in}}(x,\xi)\, \bar{u}(\xi)\, d\xi,
\end{aligned}
\end{equation}
with the transformed kernel
\begin{equation}
\tilde{K}_{\mathrm{in}}(x,\xi) = K\big(x, \phi_{\mathrm{in}}(\xi)\big)\, \left|\det \nabla \phi_{\mathrm{in}}(\xi)\right|.
\end{equation}
Therefore, an input-side transformation induces a new operator representation whose kernel incorporates two effects at once: the source location $y$ is replaced by its expression in the transformed coordinates, and the Jacobian factor compensates for the change in volume element. This is why input-side transformations modify how information is aggregated over the source domain.

\paragraph{Output-side transformation.}
Let $\psi_{\mathrm{out}}: \Omega \to \Omega$ be an output-side coordinate transformation, define transformed coordinates by $\xi = \psi_{\mathrm{out}}(x)$, and let $\phi_{\mathrm{out}} := \psi_{\mathrm{out}}^{-1}$ denote the associated inverse map. The transformed output is
\begin{equation}
\bar{v}(\xi) = v\big(\phi_{\mathrm{out}}(\xi)\big).
\end{equation}
Here the situation is different from the input-side case: the integration variable $y$ is untouched, so no Jacobian term appears. We simply evaluate the original output field at the physical point $x = \phi_{\mathrm{out}}(\xi)$ corresponding to the transformed coordinate $\xi$. Substituting this relation into the operator gives
\begin{equation}
\bar{v}(\xi)
= v\big(\phi_{\mathrm{out}}(\xi)\big)
= \int_\Omega K\big(\phi_{\mathrm{out}}(\xi), y\big)\, u(y)\, dy.
\end{equation}
This corresponds to the transformed kernel
\begin{equation}
\tilde{K}_{\mathrm{out}}(\xi,y) = K\big(\phi_{\mathrm{out}}(\xi), y\big).
\end{equation}
Hence, an output-side transformation changes the representation of the operator only through the query variable. In other words, it does not alter how source contributions are integrated, but only where the operator is sampled.

\paragraph{Summary.}
The two transformations act on complementary variables of the integral operator. Input-side transformations modify how information is aggregated from the source domain, while output-side transformations modify where the operator is evaluated. In both cases, the underlying mapping is unchanged, but the kernel representation seen by the learner is altered.
\section{Supplementary Visualization Results}
\label{app::add_vis_exp_res}

In this section, we provide comprehensive visualization results for the remaining benchmark datasets evaluated in our study. Specifically, we present qualitative comparisons between the vanilla baseline architectures (Transolver, FNO, and CNextU) and their ACT-augmented counterparts. The following figures illustrate the model predictions across various physical scenarios, including the scalar vorticity field $w$ in the \ns dataset, the activator $u$ in the \react dataset, the fluid depth $h$ in the \shallow dataset, the scalar concentration $c$ in the \actmatter dataset, and the fluid density $\rho$ in the \MHD dataset. These supplementary visual comparisons further corroborate the enhanced spatial representation and fine-grained predictive accuracy introduced by the ACT module across diverse physical dynamics.
\begin{figure}[htbp]
    \centering
    \includegraphics[width=1.0\linewidth]{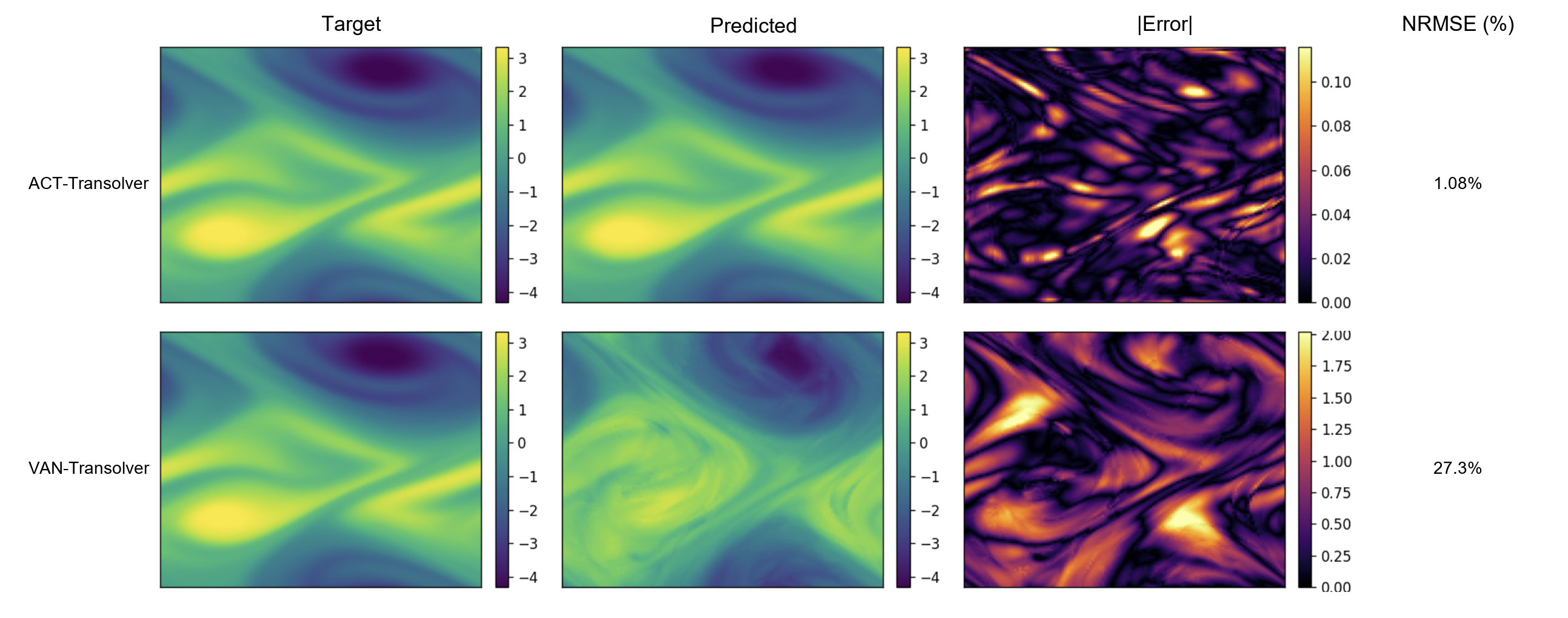}
    \caption{The scalar vorticity field $w$ on the \ns dataset comparing the vanilla Transolver and ACT-Transolver.}
\end{figure}

\begin{figure}[htbp]
    \centering
    \includegraphics[width=1.0\linewidth]{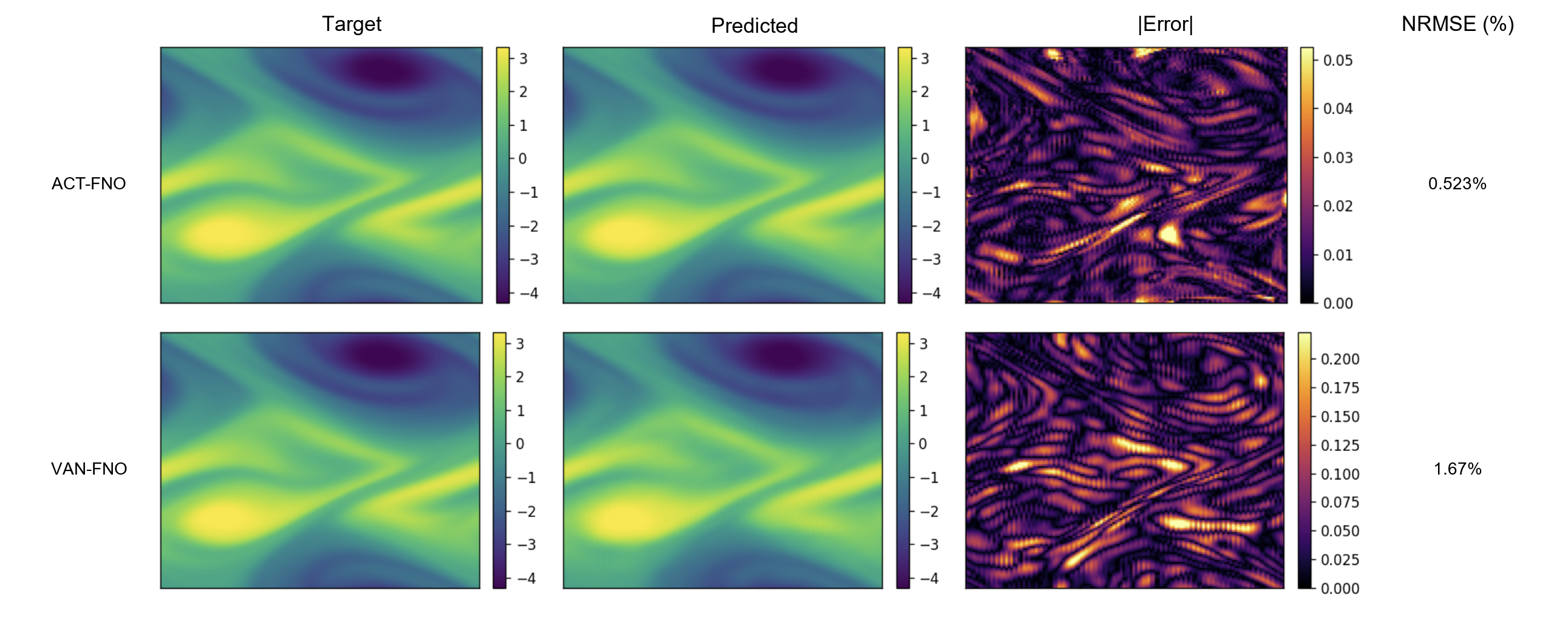}
    \caption{The scalar vorticity field $w$ on the \ns dataset comparing the vanilla FNO and ACT-FNO.}
\end{figure}

\begin{figure}[htbp]
    \centering
    \includegraphics[width=1.0\linewidth]{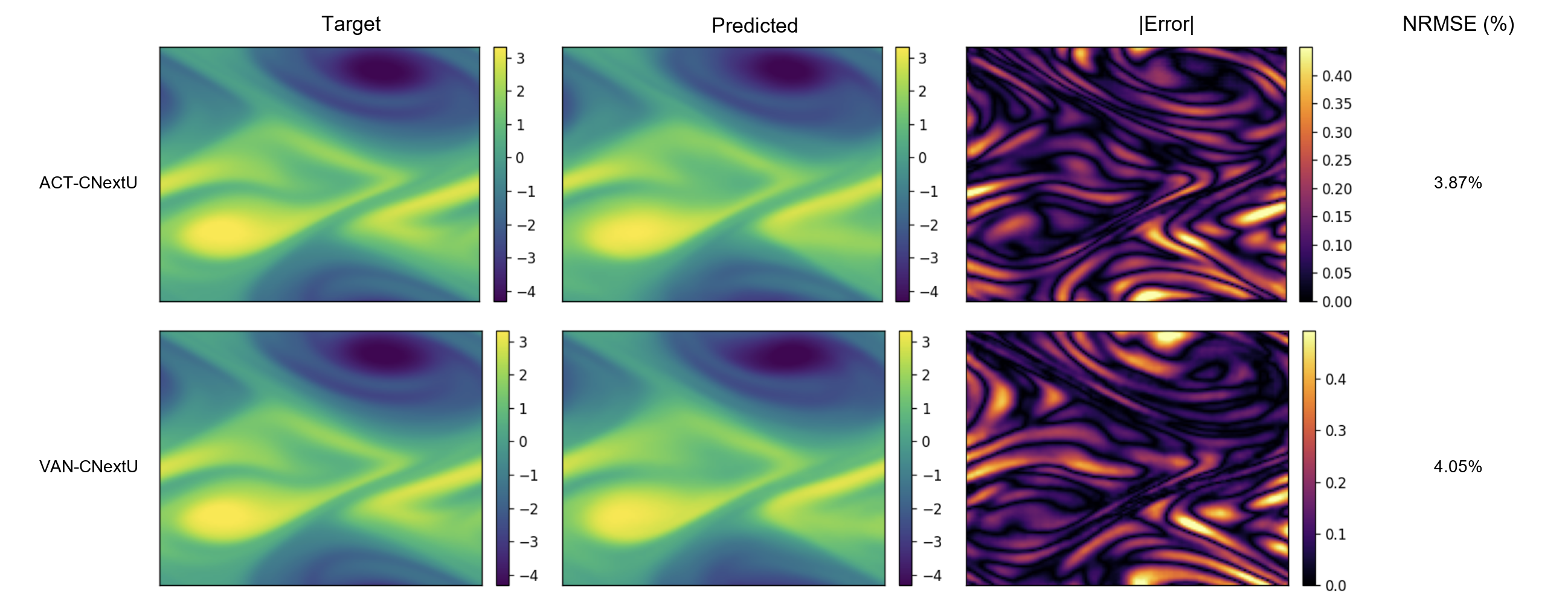}
    \caption{The scalar vorticity field $w$ on the \ns dataset comparing the vanilla CNextU and ACT-CNextU.}
\end{figure}

\begin{figure}[htbp]
    \centering
    \includegraphics[width=1.0\linewidth]{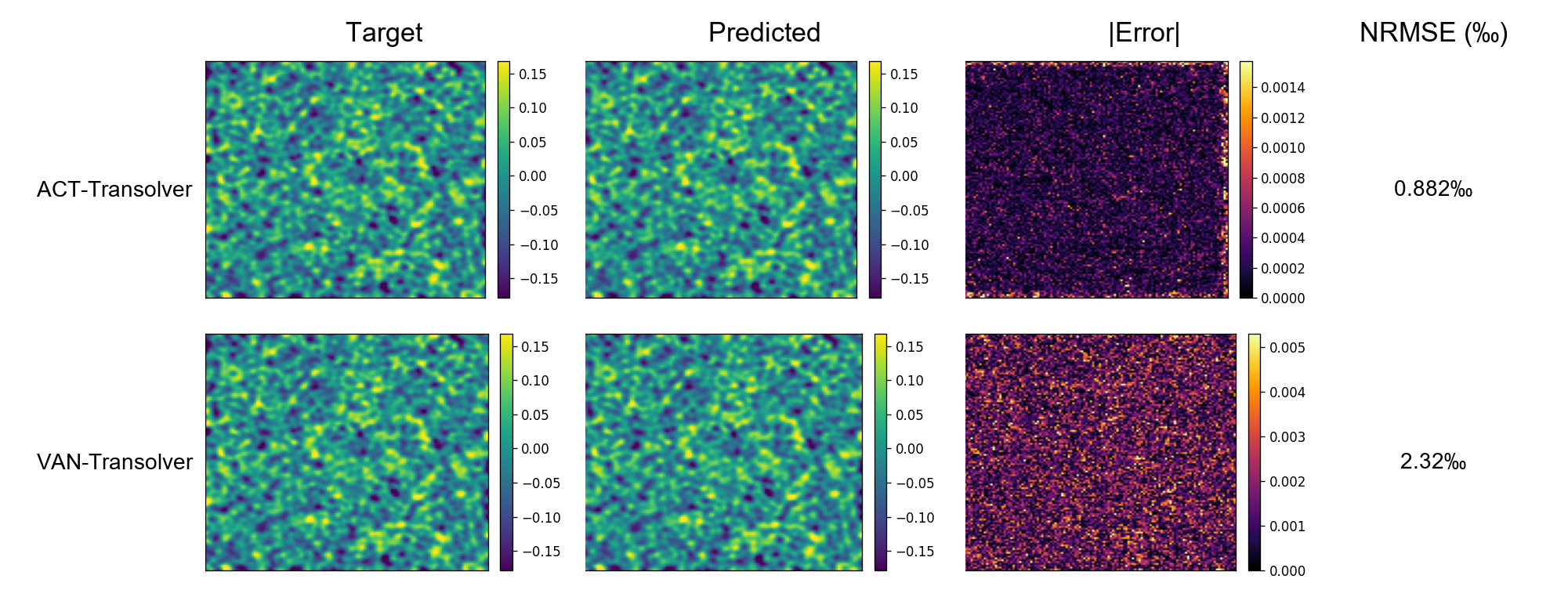}
    \caption{The activator $u$ on the \react dataset comparing the vanilla Transolver and ACT-Transolver.} 
\end{figure}

\begin{figure}[htbp]
    \centering
    \includegraphics[width=1.0\linewidth]{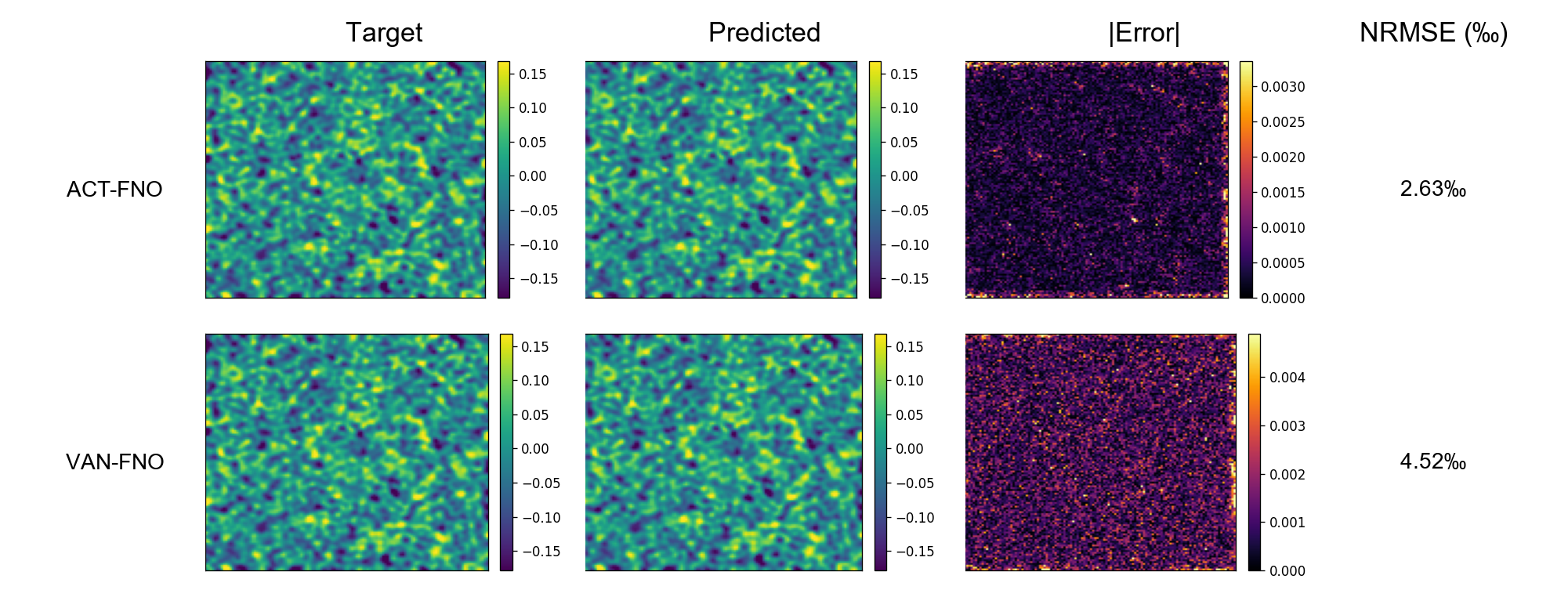}
    \caption{The activator $u$ on the \react dataset comparing the vanilla FNO and ACT-FNO.}
\end{figure}

\begin{figure}[htbp]
    \centering
    \includegraphics[width=1.0\linewidth]{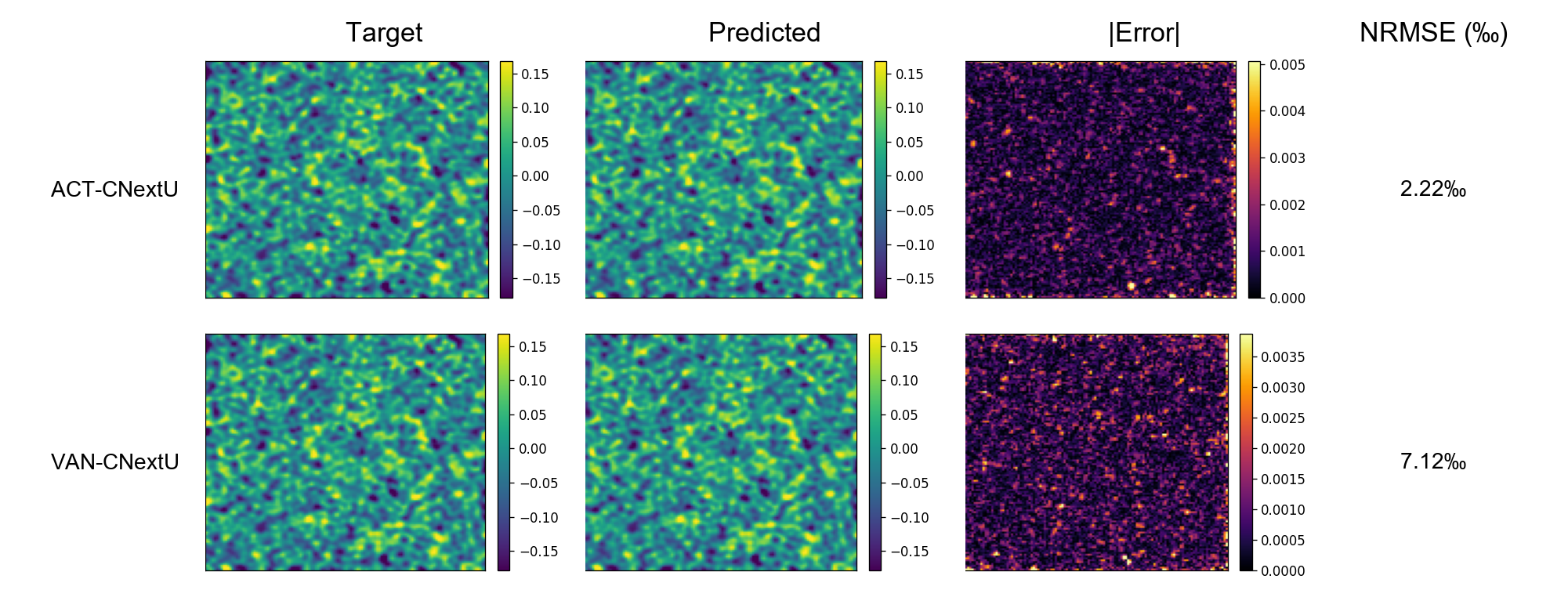}
    \caption{The activator $u$ on the \react dataset comparing the vanilla CNextU and ACT-CNextU.}
\end{figure}

\begin{figure}[htbp]
    \centering
    \includegraphics[width=1.0\linewidth]{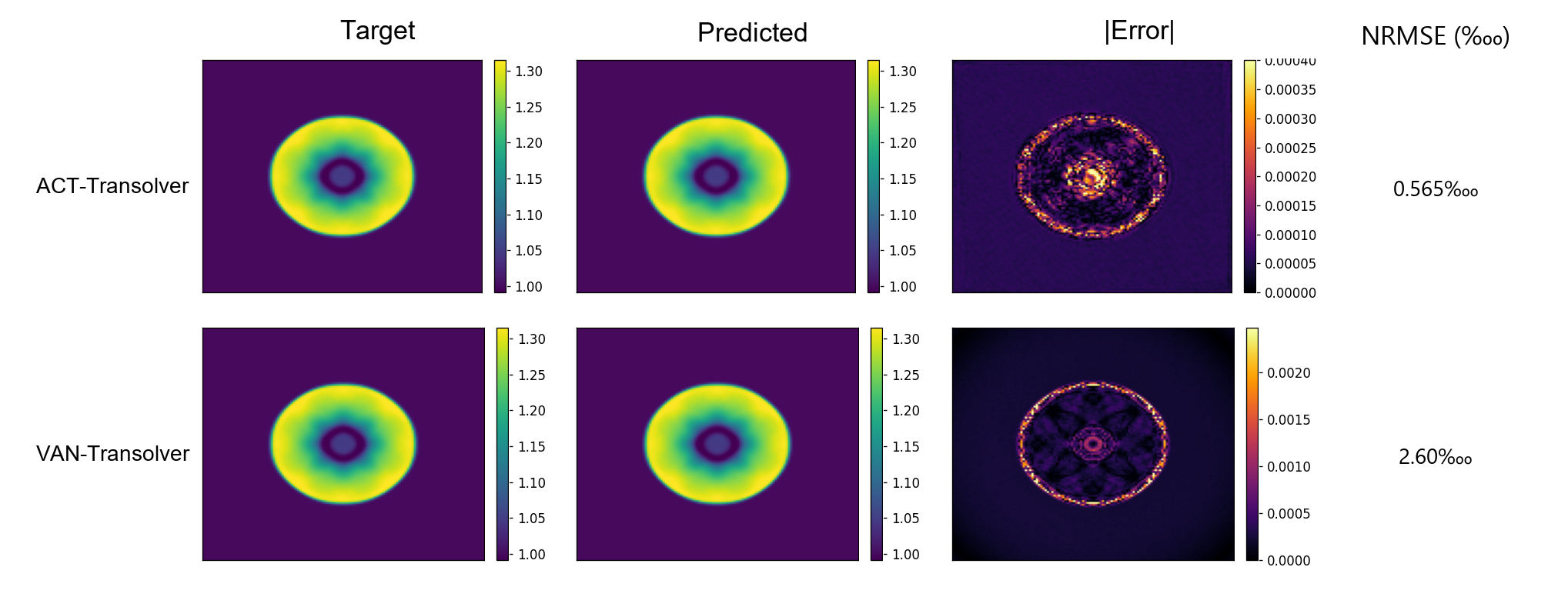}
    \caption{The fluid depth $h$ on the \shallow dataset comparing the vanilla Transolver and ACT-Transolver.}
\end{figure}

\begin{figure}[htbp]
    \centering
    \includegraphics[width=1.0\linewidth]{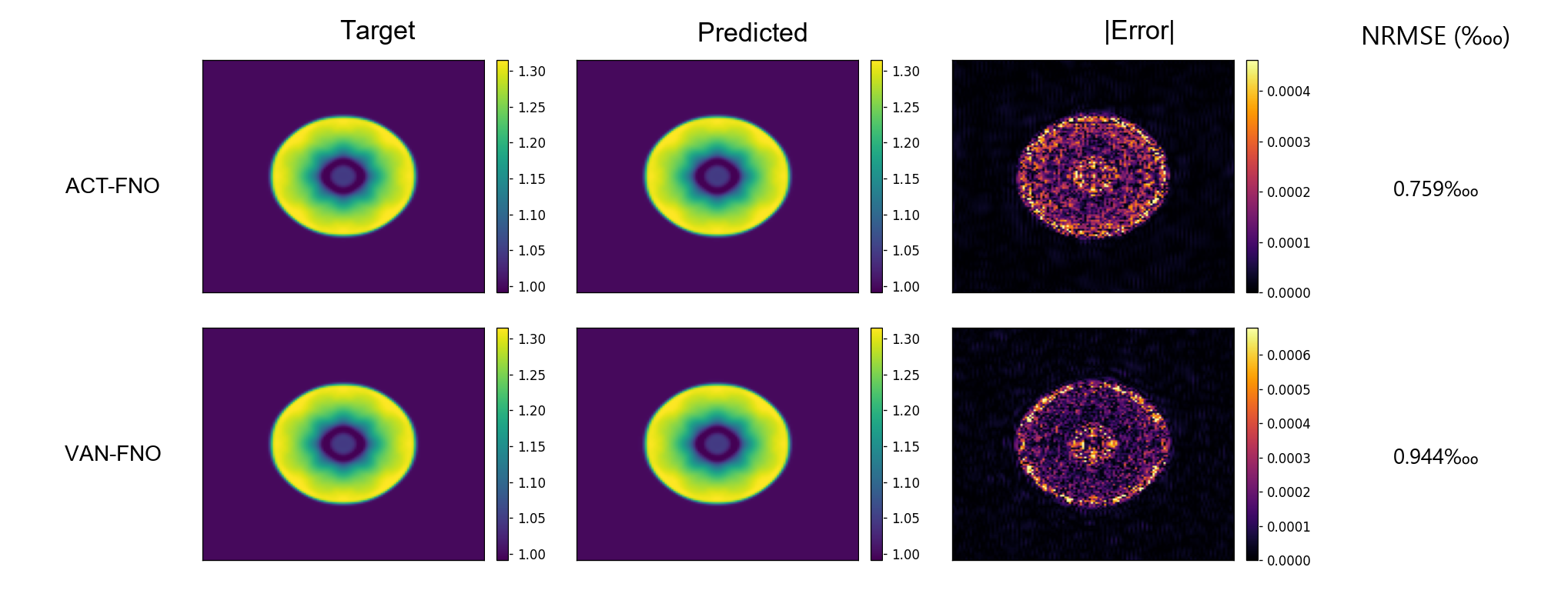}
    \caption{The fluid depth $h$ on the \shallow dataset comparing the vanilla FNO and ACT-FNO.}
\end{figure}

\begin{figure}[htbp]
    \centering
    \includegraphics[width=1.0\linewidth]{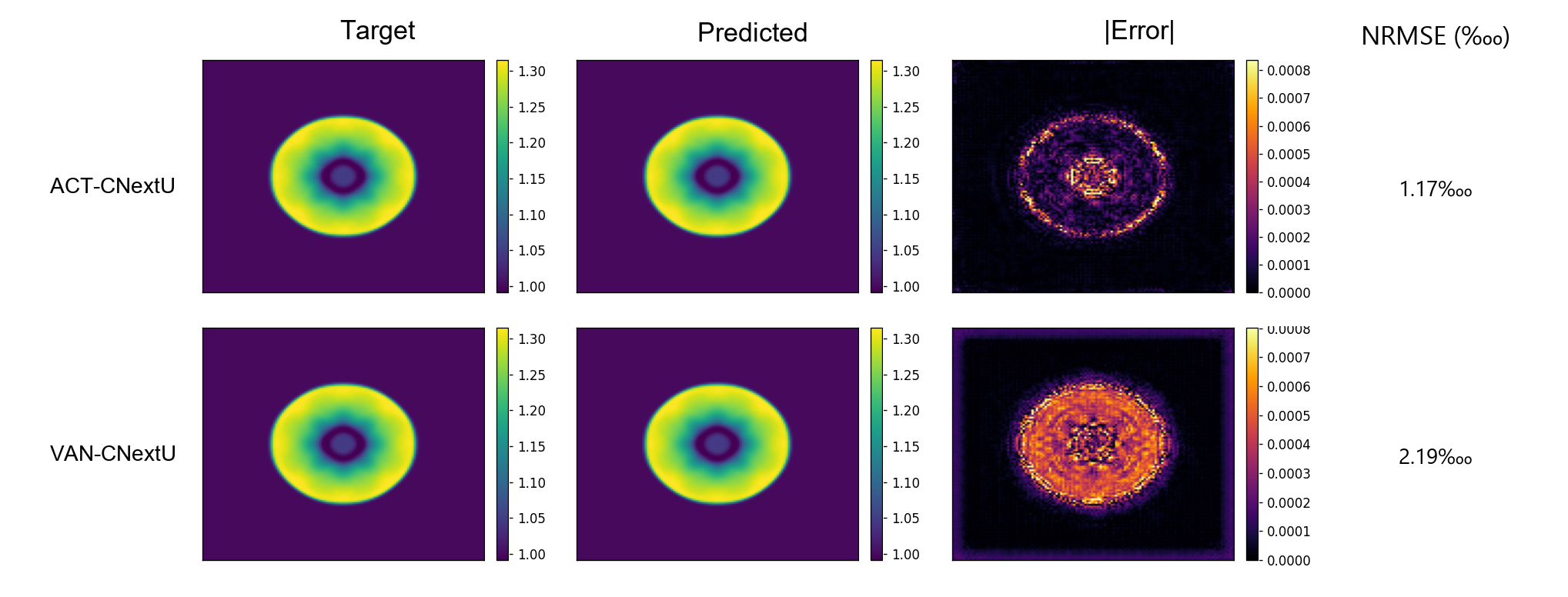}
    \caption{The fluid depth $h$ on the \shallow dataset comparing the vanilla CNextU and ACT-CNextU.}
\end{figure}

\begin{figure}[htbp]
    \centering
    \includegraphics[width=1.0\linewidth]{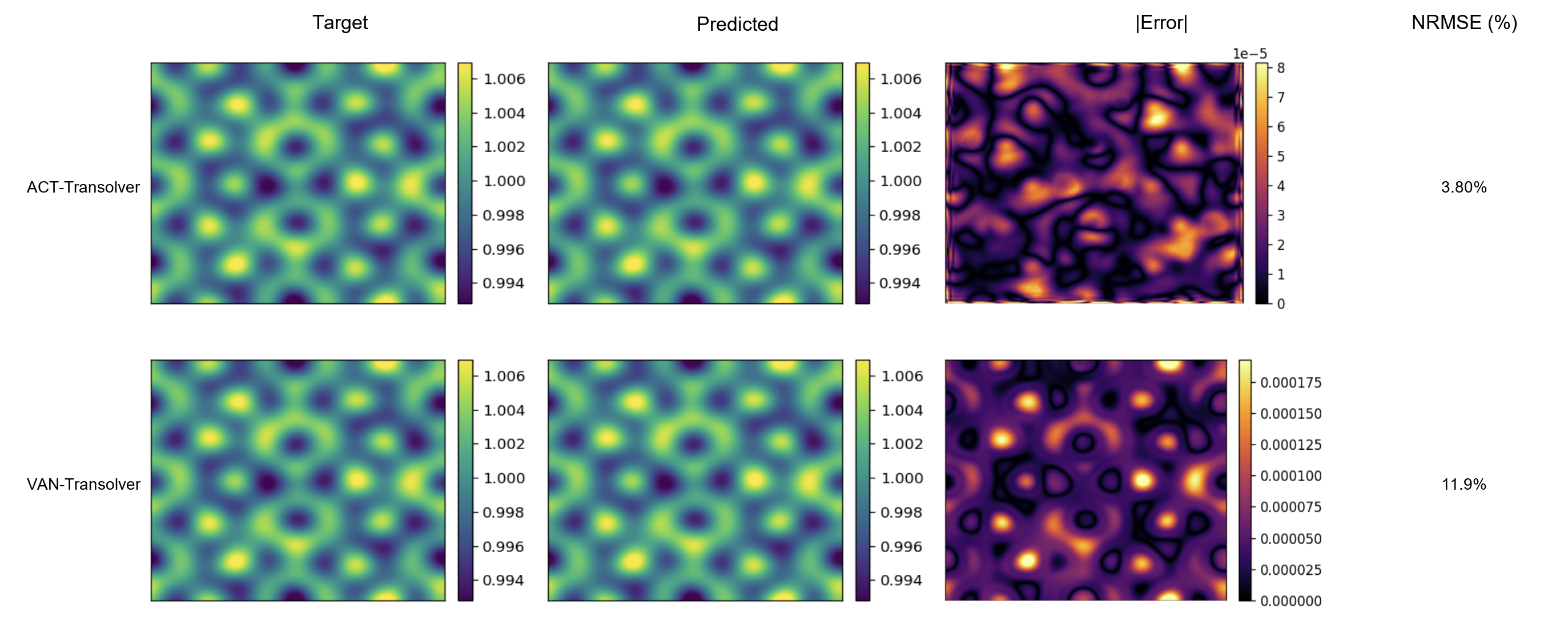}
    \caption{The scalar concentration $c$ on the \actmatter dataset comparing the vanilla Transolver and ACT-Transolver.}
\end{figure}

\begin{figure}[htbp]
    \centering
    \includegraphics[width=1.0\linewidth]{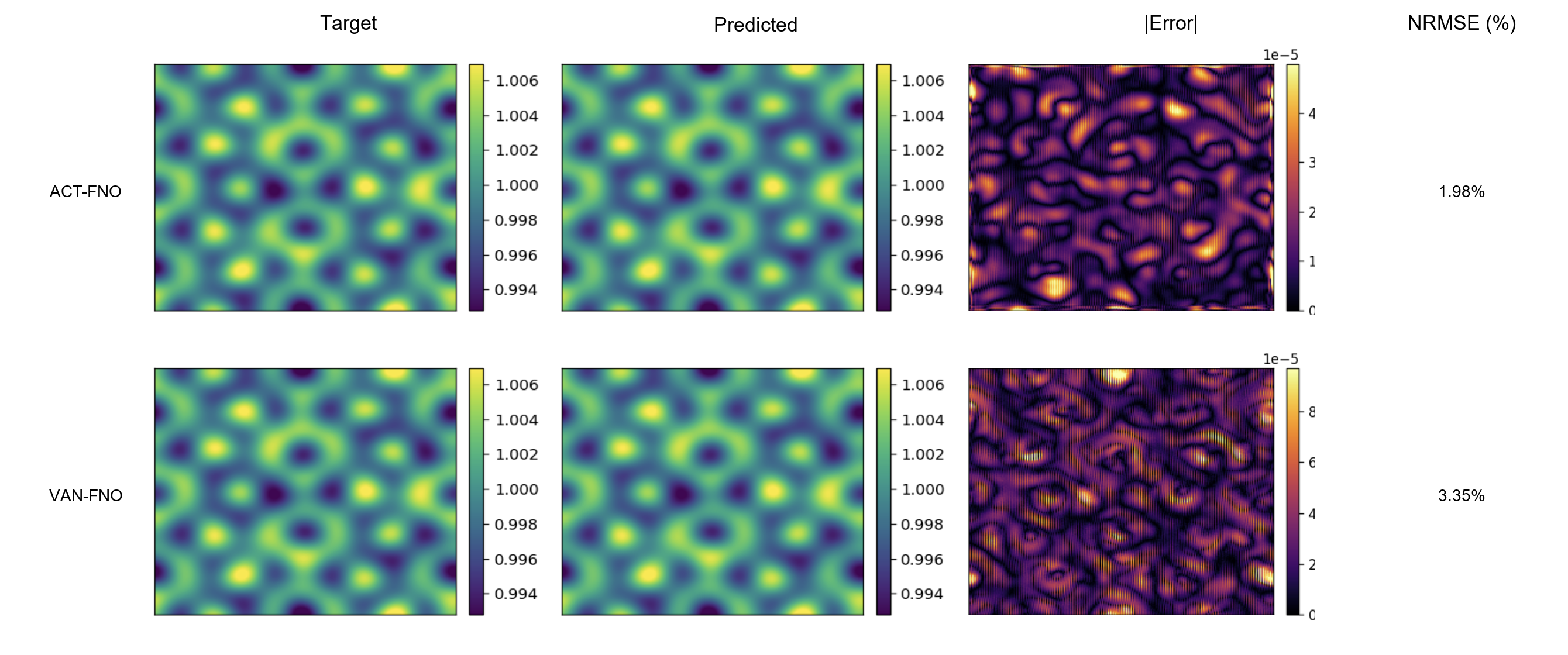}
    \caption{The scalar concentration $c$ on the \actmatter dataset comparing the vanilla FNO and ACT-FNO.}
\end{figure}

\begin{figure}[htbp]
    \centering
    \includegraphics[width=1.0\linewidth]{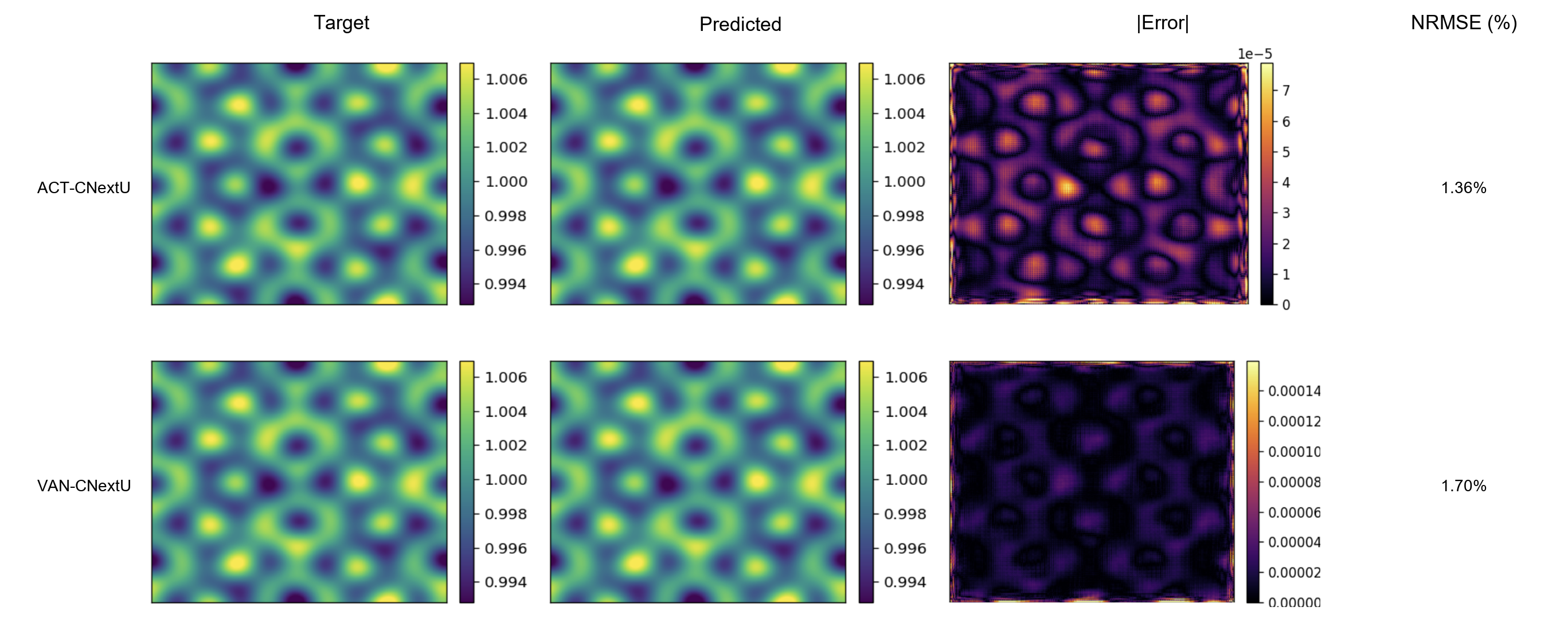}
    \caption{The scalar concentration $c$ on the \actmatter dataset comparing the vanilla CNextU and ACT-CNextU.}
\end{figure}


\begin{figure}[htbp]
    \centering
    \includegraphics[width=1.0\linewidth]{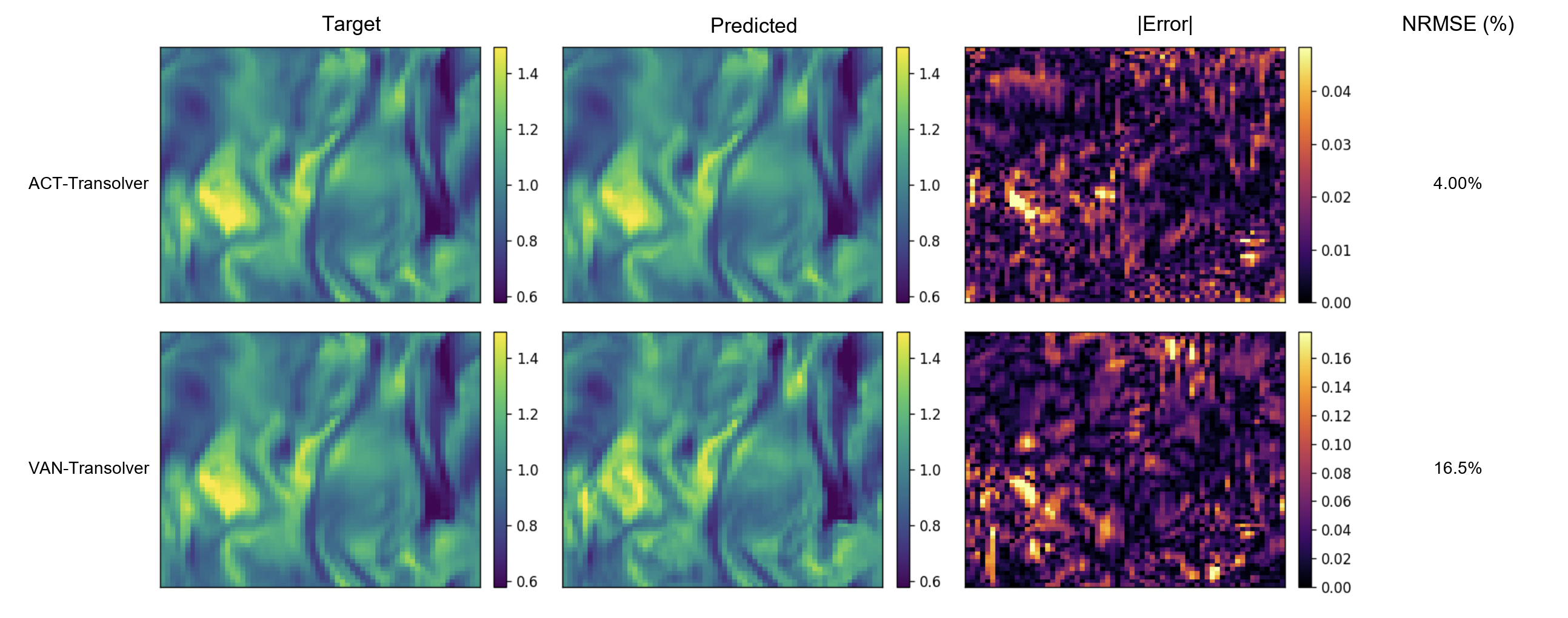}
    \caption{The fluid density $\rho$ on the \MHD dataset comparing the vanilla Transolver and ACT-Transolver.}
\end{figure}

\begin{figure}[htbp]
    \centering
    \includegraphics[width=1.0\linewidth]{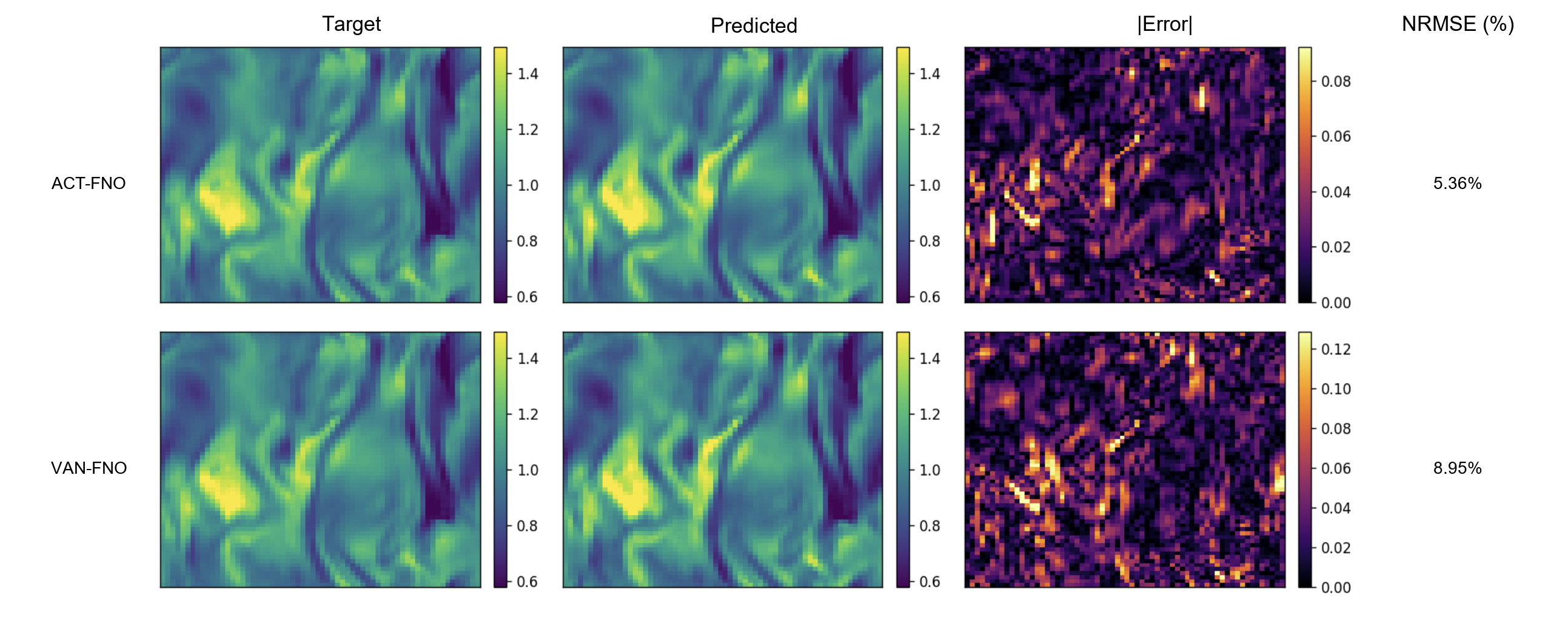}
    \caption{The fluid density $\rho$ on the \MHD dataset comparing the vanilla FNO and ACT-FNO.}
\end{figure}

\begin{figure}[htbp]
    \centering
    \includegraphics[width=1.0\linewidth]{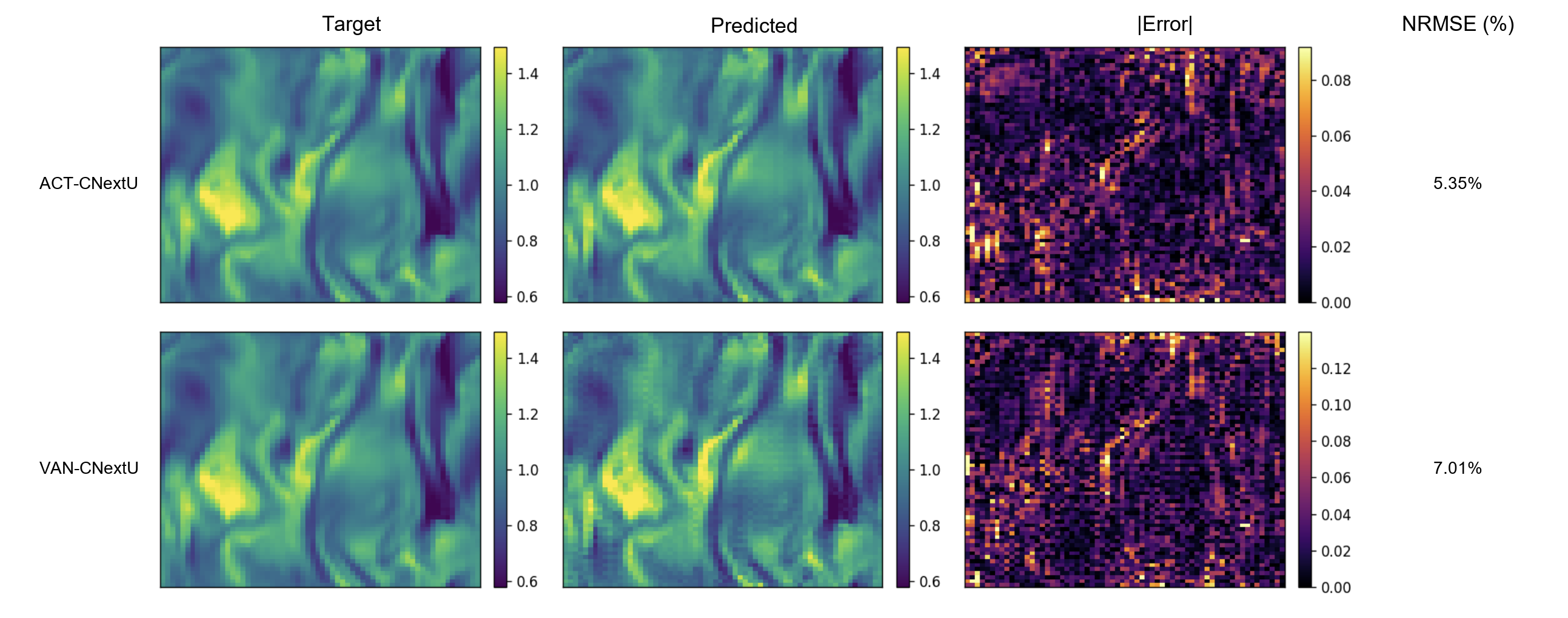}
    \caption{The fluid density $\rho$ on the \MHD dataset comparing the vanilla CNextU and ACT-CNextU.}
\end{figure}

\newpage

\FloatBarrier
\section*{NeurIPS Paper Checklist}

\begin{enumerate}

\item {\bf Claims}
    \item[] Question: Do the main claims made in the abstract and introduction accurately reflect the paper's contributions and scope?
    \item[] Answer: \answerYes{}
    \item[] Justification: The abstract and introduction clearly state the main contributions and scope.


\item {\bf Limitations}
    \item[] Question: Does the paper discuss the limitations of the work performed by the authors?
    \item[] Answer: \answerYes{}
    \item[] Justification: The paper discusses limitations in the Conclusion and Limitation section.


\item {\bf Theory assumptions and proofs}
    \item[] Question: For each theoretical result, does the paper provide the full set of assumptions and a complete (and correct) proof?
    \item[] Answer: \answerNA{}
    \item[] Justification: The paper does not present formal theorems or proofs; instead, it provides conceptual and analytical insights into how coordinate transformations affect operator representations.


    \item {\bf Experimental result reproducibility}
    \item[] Question: Does the paper fully disclose all the information needed to reproduce the main experimental results of the paper to the extent that it affects the main claims and/or conclusions of the paper (regardless of whether the code and data are provided or not)?
    \item[] Answer: \answerYes{}
    \item[] Justification: The paper provides detailed descriptions of model architectures, datasets, evaluation metrics, and training protocols sufficient to reproduce the results.
    

\item {\bf Open access to data and code}
    \item[] Question: Does the paper provide open access to the data and code, with sufficient instructions to faithfully reproduce the main experimental results, as described in supplemental material?
    \item[] Answer: \answerNo{}
    \item[] Justification: While we plan to release the code upon publication, it is not included in the current submission to preserve anonymity. All implementation details are described to support reproducibility.


\item {\bf Experimental setting/details}
    \item[] Question: Does the paper specify all the training and test details (e.g., data splits, hyperparameters, how they were chosen, type of optimizer) necessary to understand the results?
    \item[] Answer: \answerYes{}
    \item[] Justification: Training details, datasets, model configurations, and evaluation protocols are clearly specified in the main text and appendix.


\item {\bf Experiment statistical significance}
    \item[] Question: Does the paper report error bars suitably and correctly defined or other appropriate information about the statistical significance of the experiments?
    \item[] Answer: \answerNo{}
    \item[] Justification: Due to computational cost, experiments are reported without multiple runs or error bars. However, improvements are consistent across multiple datasets and architectures.
    

\item {\bf Experiments compute resources}
    \item[] Question: For each experiment, does the paper provide sufficient information on the computer resources (type of compute workers, memory, time of execution) needed to reproduce the experiments?
    \item[] Answer: \answerYes{}
    \item[] Justification: The paper specifies the hardware setup and training configurations used for experiments (see Appendix).
    
    
\item {\bf Code of ethics}
    \item[] Question: Does the research conducted in the paper conform, in every respect, with the NeurIPS Code of Ethics \url{https://neurips.cc/public/EthicsGuidelines}?
    \item[] Answer: \answerYes{}
    \item[] Justification: The research complies with the NeurIPS Code of Ethics. It involves no human subjects, sensitive data, or high-risk applications.


\item {\bf Broader impacts}
    \item[] Question: Does the paper discuss both potential positive societal impacts and negative societal impacts of the work performed?
    \item[] Answer: \answerNA{}
    \item[] Justification: This work focuses on foundational methodology for PDE learning and does not have immediate societal deployment implications.

    
\item {\bf Safeguards}
    \item[] Question: Does the paper describe safeguards that have been put in place for responsible release of data or models that have a high risk for misuse (e.g., pre-trained language models, image generators, or scraped datasets)?
    \item[] Answer: \answerNA{}
    \item[] Justification: The work does not introduce models or datasets with potential misuse risks.


\item {\bf Licenses for existing assets}
    \item[] Question: Are the creators or original owners of assets (e.g., code, data, models), used in the paper, properly credited and are the license and terms of use explicitly mentioned and properly respected?
    \item[] Answer: \answerYes{}
    \item[] Justification: All datasets and prior methods used are properly cited, and their usage follows standard academic practice.
    

\item {\bf New assets}
    \item[] Question: Are new assets introduced in the paper well documented and is the documentation provided alongside the assets?
    \item[] Answer: \answerNo{}
    \item[] Justification: The paper does not introduce new datasets. Code release is planned but not included in this submission.
    

\item {\bf Crowdsourcing and research with human subjects}
    \item[] Question: For crowdsourcing experiments and research with human subjects, does the paper include the full text of instructions given to participants and screenshots, if applicable, as well as details about compensation (if any)? 
    \item[] Answer: \answerNA{}
    \item[] Justification: The work does not involve human subjects or crowdsourcing.


\item {\bf Institutional review board (IRB) approvals or equivalent for research with human subjects}
    \item[] Question: Does the paper describe potential risks incurred by study participants, whether such risks were disclosed to the subjects, and whether Institutional Review Board (IRB) approvals (or an equivalent approval/review based on the requirements of your country or institution) were obtained?
    \item[] Answer: \answerNA{} 
    \item[] Justification: The work does not involve human subjects.
    \begin{itemize}
        \item The answer \answerNA{} means that the paper does not involve crowdsourcing nor research with human subjects.
        \item Depending on the country in which research is conducted, IRB approval (or equivalent) may be required for any human subjects research. If you obtained IRB approval, you should clearly state this in the paper. 
        \item We recognize that the procedures for this may vary significantly between institutions and locations, and we expect authors to adhere to the NeurIPS Code of Ethics and the guidelines for their institution. 
        \item For initial submissions, do not include any information that would break anonymity (if applicable), such as the institution conducting the review.
    \end{itemize}

\item {\bf Declaration of LLM usage}
    \item[] Question: Does the paper describe the usage of LLMs if it is an important, original, or non-standard component of the core methods in this research? Note that if the LLM is used only for writing, editing, or formatting purposes and does \emph{not} impact the core methodology, scientific rigor, or originality of the research, declaration is not required.
    \item[] Answer: \answerNA{}
    \item[] Justification: LLMs were not used as part of the core methodology.

\end{enumerate}

\end{document}